\documentclass[twocolumn,tighten]{aastex63}

\usepackage{amsmath}
\usepackage{comment}
\usepackage{tabularx}
\usepackage{algorithm2e}
\usepackage{float}
\usepackage{fullwidth}
\usepackage{nccmath}
\usepackage{mathtools}
\usepackage{multirow}
\usepackage{scrextend}
\usepackage{diagbox}
\usepackage{hhline}
\definecolor{yscol}{rgb}{0.8, 0.6, 1}

\newcommand{\msun}{\,{\rm M}_\odot}

\newcommand{\s}{\,{\rm s}}
\newcommand{\g}{\,{\rm g}}

\newcommand{\miest}{{\sc Miest}}
\newcommand{\tng}{{\sc IllustrisTNG}}
\newcommand{\simba}{{\sc SIMBA}}
\newcommand{\astrid}{{\sc Astrid}}
\newcommand{\eagle}{{\sc Eagle}}

\newcommand{\om}{\Omega_\mathrm{m}}
\newcommand{\sig}{\sigma_8}
\newcommand{\asnone}{A_{\rm SN1}}
\newcommand{\asntwo}{A_{\rm SN2}}
\newcommand{\aagnone}{A_{\rm AGN1}}
\newcommand{\aagntwo}{A_{\rm AGN2}}

\newcommand{\btheta}{\boldsymbol{\theta}}

\newcommand{\dd}{\mathrm{d}}

\definecolor{yscol}{rgb}{0.8, 0.6, 1}

\def\bs#1{{\boldsymbol{#1}}}

\newcommand{\tr}[1]{\textcolor{black}{#1}}

\newlength{\Oldarrayrulewidth}

\makeatletter
\newcommand{\thickhline}{%
    \noalign {\ifnum 0=`}\fi \hrule height 1pt
    \futurelet \reserved@a \@xhline
}
\newcommand{\thichline}{%
    \noalign {\ifnum 0=`}\fi \hrule height 0.8pt
    \futurelet \reserved@a \@xhline
}
\newcolumntype{"}{@{\hskip\tabcolsep\vrule width 0.8pt\hskip\tabcolsep}} 
\makeatother

\received{June 1, 2019}
\revised{January 10, 2019}
\accepted{\today}
\submitjournal{ApJ}

\graphicspath{{./}{figures/}}

\begin{document}
\title{Towards Robustness Across Cosmological Simulation Models \tng, \simba, \astrid{}, and \eagle{}}

\correspondingauthor{Yongseok Jo}
\email{yj2812@columbia.edu}
\author[0000-0003-3977-1761]{Yongseok Jo}

\author{Shy Genel}
\affiliation{Center for Computational Astrophysics, Flatiron Institute, 162 5th Avenue, New York, NY, 10010, USA}
\affiliation{Columbia Astrophysics Laboratory, Columbia University, 550 West 120th Street, New York, NY, 10027, USA}

\author[0000-0002-1080-0744]{Anirvan Sengupta}
\affiliation{Department of Physics and Astronomy, Rutgers University, Piscataway, New Jersey 08854, USA}
\affiliation{Center for Computational Quantum Physics, Flatiron Institute, New York, New York 10010, USA}
\affiliation{Center for Computational Mathematics, Flatiron Institute, New York, New York 10010, USA}

\author{Benjamin Wandelt}
\affiliation{Sorbonne Universite, CNRS, UMR 7095, Institut d’Astrophysique de Paris, 98 bis boulevard Arago, 75014 Paris, France}
\affiliation{Center for Computational Astrophysics, Flatiron Institute, 162 5th Avenue, New York, NY, 10010, USA}

\author{Rachel Somerville}
\affiliation{Center for Computational Astrophysics, Flatiron Institute, 162 5th Avenue, New York, NY, 10010, USA}

\author{Francisco Villaescusa-Navarro}
\affiliation{Center for Computational Astrophysics, Flatiron Institute, 162 5th Avenue, New York, NY, 10010, USA}
\affiliation{Department of Astrophysical Sciences, Princeton University, 4 Ivy Lane, Princeton, NJ 08544 USA}




\begin{abstract}
The rapid advancement of large-scale cosmological simulations has opened new avenues for cosmological and astrophysical research. 
However, the increasing diversity among cosmological simulation models presents a challenge to the {\it robustness}.
In this work, we develop the Model-Insensitive ESTimator (\miest{}), a machine that can {\it robustly} estimate the cosmological parameters, $\om$ and $\sig$, from neural hydrogen maps of simulation models in the CAMELS project---\tng{}, \simba{}, \astrid{}, and \eagle{}.
An estimator is considered {\it robust} if it possesses a consistent predictive power across all simulations, including those used during the training phase.
We train our machine using multiple simulation models and ensure that it only extracts common features between the models while disregarding the model-specific features. 
This allows us to develop a novel model that is capable of accurately estimating parameters across a range of simulation models, without being biased towards any particular model.
Upon the investigation of the latent space---a set of summary statistics, we find that the implementation of {\it robustness} leads to the blending of latent variables across different models, demonstrating the removal of model-specific features.
In comparison to a standard machine lacking {\it robustness}, the average performance of \miest{} on the unseen simulations during the training phase has been improved by $\sim 17\%$ for $\om$ and $38\%$ for $\sig$.
By using a machine learning approach that can extract {\it robust}\tr{, yet physical} features, we hope to improve our understanding of galaxy formation and evolution \tr{in a (subgrid) model-insensitive manner}, and ultimately, gain insight into the underlying physical processes responsible for {\it robustness}.
\end{abstract}

\keywords{methods: statistical, methods: numerical, galaxy: formation, galaxy: evolution}

\section{Introduction}
The field of cosmological simulations has undergone significant advances in recent years, driven by both the rapid development of modern technology and the increasing sophistication of computational methods \citep{pillepich2018MNRAS.473.4077P,dave2019MNRAS.486.2827D}. 
As the number of cosmological simulations continues to grow, a major challenge has emerged: the problem of calibration \citep{jo2023ApJ...944...67J}.
Furthermore, with a variety of models available, there is currently no clear consensus on which simulation method or result is the most accurate or physical. 
This issue fundamentally arises from the significant divergence in simulation outcomes, where different sub-grid models produce markedly disparate predictions, particularly for parameters on which they have not been calibrated.
This uncertainty underscores the need for more robust methods to validate and compare simulations.

Two notable projects illustrate different approaches to this issue: the AGORA project \citep{agora2014ApJS..210...14K} and the CAMELS project \citep{camels2021ApJ...915...71V} \tr{that is closely associated with the Learning the Universe (LtU) collaboration\footnote{\url{https://learning-the-universe.org}}}.
The AGORA Project endeavors to achieve convergence across different simulation codes by implementing standardized conditions and rigorously comparing outcomes. 
In contrast, the CAMELS project acknowledges and leverages the intrinsic variability among simulations, concentrating on the simulation of a diverse array of cosmological models as they intrinsically exist.

However, in the context of the AGORA project, attaining convergence proves to be a complex endeavor, frequently necessitating meticulous calibration and the acceptance of discrepancies across various simulations. Furthermore, despite its reputation as an exemplary platform for the cross-validation of diverse simulations, the resultant converged simulations do not inherently ensure physical robustness.
On the other hand, despite the extensive coverage offered by the suite of CAMELS simulations within the cosmological model space, it remains uncertain whether this space comprehensively encompasses the true structure of the Universe. 
These competing methodologies give rise to a pivotal question: ``Should we prioritize the execution of a broader array of diverse simulations, or should our efforts be concentrated on the development of a convergent model capable of replicating a maximal number of observations?''

Concurrently, machine learning techniques are progressively being incorporated into cosmological simulations through various approaches, including the acceleration and enhancement of the simulation process \citep{jo2019MNRAS.489.3565J,he2019PNAS..11613825H,nguyen2023arXiv230805145N,jamieson2023ApJ...952..145J}, as well as the estimation of cosmological and astrophysical parameters \citep{shao2023ApJ...956..149S,natali2023arXiv231015234D,hahn2023arXiv231015246H,max2024ApJ...968...11L}.
These endeavors have demonstrated significant progress and have become an essential component of cosmological simulations.
Nevertheless, investigations that focus on the variety of cosmological simulations and their corresponding model spaces are relatively absent.
In particular, the problem of {\it robustness} has emerged within the context of estimations.
{\it Robustness} here refers to a phenomenon in which an estimator exhibits consistent performance across different models including those that are not included during the training phase.
Despite the efforts with regards to the robustness of cosmological simulations \citep{shao2022arXiv220906843S,rojas2023ApJ...954..125E,natali2023ApJ...952...69D,domingo2022ApJ...937..115V}, these attempts are limited by the dependence on manual identification of reliable physical quantities for parameter estimation.

In the meantime, the machine learning community is endeavoring to establish a rigorous methodology for achieving robustness, often referred to as domain generalization \citep{li10.1007/978-3-030-01267-0_38,muandet10.5555/3042817.3042820,finn2018metalearninguniversalitydeeprepresentations,blanchard10.5555/3546258.3546260,zhao2019learninginvariantrepresentationdomain,wang9782500,haung10.1007/978-3-030-58536-5_8,zhang2023adversarialstyleaugmentationdomain,zhou9847099}.
Domain generalization aims to train models that can generalize effectively to new, unseen domains without requiring access to data from those domains during training. 
In contrast to domain adaptation, which permits models to access target domain data (either labeled or unlabeled) to adjust their learning, domain generalization prohibits any such access. 
This constitutes a formidable yet highly pragmatic challenge, particularly in real-world scenarios where new domains may emerge continuously. 
Domain generalization usually includes some pivotal techniques such as domain shift, invariant feature learning, meta-learning, regularization, ensemble learning, etc.

The recent work in astrophysics has adopted domain shift to infer cosmological parameters from observational data, utilizing a machine trained on multiple simulations \citep{lee2024arXiv240902256L}.
The domain shift adopted in \citet{lee2024arXiv240902256L} works in a way that introduces certain shifts in parameter space so that the parameters can be tuned to the true values.
However, a significant limitation in this tuning process is the reliance on SDSS-calibrated simulations for model calibration.
\tr{Similarly, \citep{andrianomena2025Ap&SS.370...14A} presents a method to address covariate shift between datasets of the same observable with different distributions, improving the generalizability of neural networks for cosmological inference. 
Using HI maps from IllustrisTNG and SIMBA, the authors employ adversarial training and optimal transport to adapt a pre-trained neural network for out-of-distribution data without requiring labels.
However, this does not constitute true generalization in principle, as the model has indirectly learned about the target dataset through domain adaptation.}


\tr{As part of LtU collaboration that aims to reconstruct cosmological parameters, initial conditions, and astrophysical processes from upcoming observational data by leveraging machine learning to accelerate forward modeling,}
our primary objective is to address the robustness of multiple cosmological simulations using domain generalization in an adversarial fashion.
In the line of efforts of having diverse universe, the suite of the CAMELS simulations now possesses several different cosmological simulation models, including \tng{}, \simba{}, \astrid{}, and \eagle{} that have reached $z=0$ with 1000 variations of cosmological and astrophysical parameters.
Previous research has demonstrated that the different cosmological models exhibit unique signatures, especially in temperature maps and gas density maps \citep{camels2021ApJ...915...71V,camelsmaps2022ApJS..259...61V,tillman2023AJ....166..228T,busillo2023MNRAS.525.6191B,medlock2024ApJ...967...32M}, which subsequently influence the cross-estimation of cosmological parameters with a significant bias across different cosmological models; {\it i.e., they are not robust} \citep[e.g.,][]{paco2021arXiv210909747V,paco2021arXiv210910360V,ni2023ApJ...959..136N}. 
Amongst these maps, this work focuses on neutral hydrogen (HI) maps. 
Neutral hydrogen is a key tracer of dark matter and large-scale structure in the universe, making it a primary focus for several ongoing and future cosmological surveys, including major projects like the Square Kilometer Array (SKA), Hydrogen Epoch of Reionization Array (HERA), Low Frequency Array (LOFAR), the Vera C. Rubin Observatory's Legacy Survey of Space and Time (LSST), the Nancy Grace Roman Space Telescope, SPHEREx, and Euclid. 
Moreover, HI maps have been employed as tracers of cosmology within the realm of machine learning applications. \citep[e.g.,][]{hassan2022ApJ...937...83H}.


With HI maps, we build a Model-Insensitive ESTimator (\miest{}) that estimates cosmological parameters in a {\it robust} manner that consistently performs on different cosmological simulation models.
\miest{} processes HI maps using a convolutional neural network (CNN) and compress them into the latent variables, inspired by the Information Bottleneck (IB) method \citep{tishby1999proceedings}.
We seek to have these latent variables have limited information on the the type of simulation that generated the original data.
While \citet{alemi2017deep} has provided a variation approach using a neural network to IB, we choose a slightly different objective function, which utilizes de-classification.
In our approach, the latent variables are trained to get rid of classifiable information that is intrinsic to each simulation model, achieving robustness in \miest{}.
In this end , we aim at providing new insights into the configuration of simulations within the cosmological model space, thereby paving the way for more reliable comparisons and a clearer understanding of simulation results. 
Ultimately, we anticipate that our findings could suggest new directions for future cosmological simulations, helping to shape the field's evolution toward greater accuracy and consistency.

The structure of this paper is as follows: 
In Sec. \ref{sec:method_camels}, we introduce the CAMELS simulations and the descriptions of \tng, \simba, \astrid, and \eagle{} that we adopt.
In Sec. \ref{sec:method_neural_network}, we describe the structure and training procedure of \miest{} in detail.
Sec. \ref{sec:method_dimension_reduction} summarizes the dimensionality reduction tools adopted to visualize the latent space.
In Sec. \ref{sec:latent_space}, we investigate the properties of the latent space in response to implementation of robustness.
In Sec. \ref{sec:performance}, we provides the quantitative analysis on the improved performance of the \miest{} with robustness.
In Sec. \ref{sec:discussion}, we discuss the definition of robustness, stability and convergence of \miest{}, and comparisons between any two simulations.
Sec. \ref{sec:summary} provides the summary of the \miest{}.

\section{Methodology}
\label{sec:method}
\subsection{Cosmological Simulations}
\label{sec:method_camels}
Cosmology and Astrophysics with MachinE Learning Simulations (CAMELS)\footnote{\url{https://www.camel-simulations.org}} is a suite of thousands of cosmological simulations run with various cosmological models such as \tng, \simba, \astrid{}, and \eagle{} \citep{camels2021ApJ...915...71V,li10.1007/978-3-030-01267-0_38}. 
Each simulation contains $256^3$ dark matter particles of mass $6.49 \times 10^7 (\Omega_m - \Omega_b)/0.251 h^{-1} \msun$ and $256^3$ gas cells with an initial mass of $1.27 \times 10^7 h^{-1}\msun$ in a periodic box of a comoving volume of $(25\,h^{-1}\mathrm{Mpc})^3$.
The cosmological and astrophysical parameters of interest are $\Omega_\mathrm{m}$, $\sigma_8$, $A_\mathrm{SN1}$, $A_\mathrm{SN2}$, $A_\mathrm{AGN1}$, and $A_\mathrm{AGN2}$ (refer to Sec. 3 in \citet{camels2021ApJ...915...71V} for details).
In this study, We adopt the LH suite for the training dataset.
The 1000 simulations of the LH set are run with $\Omega_\mathrm{m} \in [0.1, 0.5]$, $\sigma_8 \in [0.6,1.0]$, $A_\mathrm{SN1}\in [0.25, 4.0]$, $A_\mathrm{SN2}\in [0.5,2.0]$, $A_\mathrm{AGN1}\in [0.25,4.0]$, and $A_\mathrm{AGN2}\in [0.5,2.0]$ arranged in a latin hypercube.
The cosmological parameters are fixed in all simulations as $\Omega_\mathrm{b} = 0.049,\, h=0.6711,\, n_\mathrm{s}=0.9624,\, M_\nu =0.0eV,\, w = -1$ and $\Omega_\mathrm{K}=0$. 

\subsubsection{IllustridTNG}
The IllustrisTNG model utilizes the {\sc Arepo} code \citep{springel2010ARA&A..48..391S,weinberger2020ApJS..248...32W} to solve the coupled equations of gravity using TreePM and magnetohydrodynamics using a Voronoi moving mesh approach, and includes a comprehensive set of baryonic physics modules \citep[for a full description]{weinberger2017MNRAS.465.3291W,pillepich2018MNRAS.473.4077P}.
The subgrid models cover radiative cooling and heating \citep{katz1996ApJS..105...19K,wiersma2009MNRAS.399..574W}, hydrogen self-shielding \citep{rahmati2013MNRAS.430.2427R}, star formation, and stellar evolution \citep{springel2003MNRAS.339..289S, vogelsberger2013MNRAS.436.3031V}. 

In addition, galactic winds from stellar feedback are implemented kinetically through temporarily hydrodynamically decoupled particles, ejected stochastically and isotropically from star-forming gas, based on \citet{springel2003MNRAS.339..289S} with 10\% thermal energy addition. 
Wind speed and energy injection rate per unit star formation depend on gas conditions (metallicity and dark matter velocity dispersion), redshift, and two global normalization parameters.
Here, the energy injection rate and the wind speed are modulated by the parameters $A_\mathrm{SN1}$ and $A_\mathrm{SN2}$, respectively.

SMBH particles with initial mass $M_\mathrm{seed} = 8 \times 10^5 \msun/h$ are seeded in halos with $M_\mathrm{FoF} > 5 \times 10^{10} \msun/h$ \citep[for details of the SMBH model]{weinberger2017MNRAS.465.3291W}. 
SMBH particles are repositioned at the potential minimum and merge if within each other's feedback spheres. Gas accretion follows the spherical \citet{bondi1952MNRAS.112..195B} with the Eddington cap.
SMBH feedback employs thermal, kinetic, and radiative modes. 
Thermal and kinetic modes operate separately, distinguished by the Eddington ratio of the SMBH's accretion rate. 
The transition from high to low mode generally occurs around $M_\mathrm{SMBH} \sim 10^8\msun$. 
The high accretion rate thermal mode injects thermal energy into a feedback sphere with an energy conversion efficiency of 0.02.
The astrophysical parameters $\aagnone$ and $\aagntwo$ govern energy injection in the kinetic feedback mode, dependent upon the accretion rate and the total energy allocation in the low accretion mode.

\subsubsection{SIMBA}
The SIMBA model utilizes the N-body/hydrodynamics code {\sc GIZMO} \citep{hopkins2015MNRAS.450...53H} with the ``Meshless Finite Mass'' hydrodynamics solver \citep[for a full description]{dave2019MNRAS.486.2827D}.
Gravitational forces are computed using the modified TreePM algorithm of the GADGET-III code \citep{springel2005MNRAS.364.1105S}, including adaptive gravitational softenings for the gas, stellar, and dark matter components.
The subgrid models include radiative cooling and photoionization \citep{haardt2012ApJ...746..125H,smith2017MNRAS.466.2217S}, hydrogen self-shielding \citep{rahmati2013MNRAS.430.2427R}, star formation and feedback and evolution \citep{krumholz2011ApJ...729...36K, dave2016MNRAS.462.3265D}.

Stellar feedback drives galactic winds that are kinetically implemented through hydrodynamically decoupled, two-phase, metal-enriched outflows. In these outflows, 30\% of the wind particles are heated to a temperature determined by subtracting the wind kinetic energy from the SNe energy. Star-forming gas elements are ejected stochastically based on predetermined values of the mass loading factor and wind speed. The mass loading factor is derived from the FIRE ``zoom-in'' simulations \citep{hopkins2014MNRAS.445..581H} and adjusts according to the galaxy's stellar mass, in line with \citet{angles2017MNRAS.470.4698A}. 
In CAMELS, $\asnone$ is introduced to regulate the overall normalization of the mass loading factor, while $\asntwo$ adjusts the wind speed normalization for varying parameters.

The black hole model in \simba{} includes gravitational
torque accretion and kinetic feedback \citep{angles2017MNRAS.464.2840A}. 
Galaxies with stellar mass $M_\star > 10^{9.5} \msun$ are seeded with black holes of initial mass $M_\mathrm{seed} = 10^4 \msun/h$ using the on-the-fly FoF method, based on the stellar mass threshold for seeding.
Black hole particles are relocated to the potential minimum's position within the FoF host group if it lies within a distance of $<4\times R_0$, where $R_0$ represents the size of the black hole kernel that encloses the nearest 256 gas elements. 
Black holes within $R_0$ of each other merge instantaneously if their relative velocity is less than three times their mutual escape velocity.
The accretion processes governing black hole growth encompass the accretion of cold gas as delineated by \citet{hopkins2011MNRAS.415.1027H}, as well as the accretion of hot gas described by \citet{bondi1952MNRAS.112..195B}.
Black hole feedback consists of high mass loading outflows in the radiative ``QSO'' mode and lower mass loading but faster outflows at low Eddington ratios in the jet mode, both of which eject gas elements bipolarly.
Black holes with $M_\mathrm{BH} > 10^{7.5} \msun$ that accrete below an Eddington ratio of 0.2 are switched to the jet mode.
$\aagnone$ modulates the total momentum flux of both modes, while $\aagntwo$ controls the maximum jet speed for the jet mode \citep[for a full review of the black hole model]{dave2019MNRAS.486.2827D}.

\subsubsection{ASTRID}
The ASTRID model utilizes a new
version of the MP-Gadget simulation code, a massively
scalable version of the cosmological structure formation code
Gadget-3 \citep{springel2005MNRAS.364.1105S}, to solve gravity with TreePM and hydrodynamics with a
smoothed particle hydrodynamics method (full review \citet{bird2022MNRAS.512.3703B} and \citet{ni2023ApJ...959..136N}). 
The subgrid model includes radiative cooling and photoionization heating \citep{katz1996ApJS..105...19K}, metal line
cooling with tracing the gas and stellar metallicities \citep{vogelsberger2014MNRAS.444.1518V}, a spatially uniform ionizing
background \citep{faucher2020MNRAS.493.1614F}, hydrogen self-shielding \citep{rahmati2013MNRAS.430.2427R}, and star formation \citep{springel2003MNRAS.339..289S}.

Winds driven by stellar feedback within galaxies are kinetically implemented through particles that are temporarily decoupled from hydrodynamics. 
These winds originate from newly formed star particles, which randomly select gas particles within their SPH smoothing length to transform into wind particles. 
A particle becomes recoupled after its density decreases by a factor of 10. 
Wind particles do not engage in or generate pressure but do receive mass return, undergo cooling, and participate in density estimations. 
In the CAMELS suite, $\asnone$ controls the total energy injection rate per star formation, while $\asntwo$ regulates the speed of the SN wind.

The SMBH model adopted the BH seeding and dynamics prescriptions applied in {\sc BlueTides} \citep{feng2016MNRAS.455.2778F}. 
BHs with initial mass $M_\mathrm{seed} = 5\times10^5 \msun/h$ are seeded in halos with $5\times 10^{10}\msun/h$ by means of the on-the-fly friends-of-friends (FOF) halo finding. 
BH particles are repositioned to the location of the local potential minimum at each active time step, and two BHs located within twice the gravitational softening length of each other are
instantaneously merged \citep{ni2022ApJ...940L..49N}. 
BH accretion rates follow the Bondi–Hoyle–Lyttleton-like prescription \citep{matteo2005Natur.433..604D}.
Super-Eddington accretion periods are allowed, but are capped at twice the Eddington rate.
BH bolometric luminosity is proportional to the accretion rate with a 0.1 mass-to-light conversion efficiency \citep{shakura1973A&A....24..337S}.
SMBH feedback uses a two-mode approach: thermal and kinetic, based on the Eddington ratio of the current BH accretion rate.
The kinetic mode is activated contingent upon an Eddington threshold for black holes with $M_\mathrm{BH}=5\times10^8 \msun/h$ with an Eddington threshold cap of 0.05.
$\aagnone$ and $\aagntwo$ modulate the kinetic and thermal feedback efficiency, respectively, with $\aagnone$ retaining the same physical meaning as in the TNG suite.

\subsubsection{EAGLE}
The \eagle{} model utilizes the smoothed particle hydrodynamics and gravity code Swift. 
Swift is a parallel, open-source, versatile and modular code, with a range of hydrodynamics solvers, gravity solvers, and sub-grid models for galaxy formation.
In the CAMELS suite, the SPHENIX flavour of SPH is adopted, coupled with a modified version of the Evolution and Assembly of GaLaxies and their Environments (\eagle{}) subgrid model for galaxy formation and evolution (see \citet{schaye2015MNRAS.446..521S} and \citet{crain2015MNRAS.450.1937C}). 
This includes element-by-element radiative cooling and heating rates from \citet{poloeckinger2020MNRAS.497.4857P}, star formation \citep{schaye2008MNRAS.383.1210S}, stellar evolution and enrichment from \citet{wiersma2009MNRAS.399..574W}, and single thermal-mode feedback from massive stars and accreting AGN \citep[see][]{booth2009MNRAS.398...53B,dalla2012MNRAS.426..140D,rosas-guevara2015MNRAS.454.1038R}.

\subsubsection{HI Maps}
In this study, we employ projected two-dimensional neutral hydrogen (HI) density maps as input to allow the machine to estimate the cosmological parameters, utilizing the CAMELS Multifield Dataset \citep{camelsmaps2022ApJS..259...61V}.
The HI field represents the spatial distribution of the neutral hydrogen density. 
To create 2D maps, it is necessary to read the locations and neutral hydrogen masses of all gas particles within a given simulation. 
The neutral hydrogen mass of each gas particle is determined using the self-shielding fitting formula outlined in \citet{rahmati2013MNRAS.430.2427R}, while star-forming particles are assumed to be completely self-shielded from external radiation, as described in \citet{villaescusa2018ApJ...866..135V}. 
The value stored in each pixel is the total neutral hydrogen density from all gas particles that contribute to that specific location, or more explicitly
\begin{equation}
    \frac{1}{Q}\sum_{i}M_{\mathrm{HI},i}\int_{x}W_{g,i}(\bs{x}-\bs{r}_{g,i})\dd \bs{x},
    \label{eq:HI}
\end{equation}
where $W_{g,i}$ is the window function, $M_{\mathrm{HI},i}$ is the neutral hydrogen mass, and $\bs{r}_{g,i}$ is the position of gas particle $i$. 
The sum and integral are over all gas particles and the area of the pixel, respectively. 
$Q$ denotes the pixel area in 2D maps.

Two-dimensional maps are constructed by initially extracting the spatial coordinates and attributes of a specified field from the simulation data. 
Each simulation comprises 15 slices: five in each of the XY, XZ, and YZ planes, respectively. 
Particles located within a slice of the volume $(25 \times 25 \times 5)(h^{-1}\mathrm{Mpc})^3$ are selected, such that slices oriented in the same projection direction are non-overlapping. 
The particle radius $R$ is determined by calculating the distance to the 32nd nearest gas particle, and the window function of the particles, $W_{2D} (r,\theta)$, is mapped onto a two-dimensional plane, where
\begin{equation}
W_\mathrm{3D}(r,\theta,z) = 
\left\{ 
\begin{array}{cc} 
\frac{3}{4\pi R^3} & \mbox{if}  |r|<R \\
0 & \mbox{otherwise}
\end{array}\right.,
\end{equation}
and $W_\mathrm{2D}(r,\theta)=\int_{z}W_\mathrm{3D}(r,\theta,z)\dd z$.
Lastly, the attributes of the particles are superimposed on a grid of $256 \times 256$ using Eq.~\ref{eq:HI}. 
This procedure involves uniformly sampling 1000 tracers within each particle radius, subsequently assigning each tracer to the corresponding pixel. 
This numerical approximation method asymptotically approaches the precise outcome as the number of tracers increases.
In summary, the maps are generated at a redshift of z = 0 and consist of $256 \times 256$ pixels spanning an area of $25 \times 25 (h^{-1} \mathrm{Mpc})^2$. 
A total of 15,000 maps are produced for each simulation suite, with 15 maps per simulation in 1000 different simulations.

The correlation of the HI maps with cosmological parameters---$\om$ and $\sig$---overall resembles the correlation between the large-scale structures of dark matter and cosmological parameters.
For instance, larger $\om$ results in more pronounced structures as more matter is available for gravitational collapse, while a higher $\sig$ leads to more tightly clustered, clumpy structures due to enhanced fluctuations.
However, a notable difference from dark matter is that neural hydrogen is more sensitive to subgrid models or astrophysical phenomena such AGNs and supernovae.
Subsequently, the disparity in subgrid models of cosmological simulations are imprinted on HI maps through the variation of the cosmological parameters.
For example, in the case of AGN feedback, increasing the matter content in the universe (high $\om$) results in the formation of dense clusters and galaxies, which facilitates the growth of supermassive black holes and subsequently leads to a greater fraction of ionized diffuse gas in the universe due to the jet feedback from these black holes.
This can produce effects comparable to the increase in the AGN parameters. 
Consequently, the distinct subgrid models employed in cosmological simulations can significantly influence the notable disparity in HI maps across cosmological simulation models (refer to Sec.~\ref{sec:discussion_difference_in_simulations} and see Fig.~\ref{fig:hi_maps_om} and \ref{fig:hi_maps_sig} for the detailed discussion).

\begin{figure}
    \centering
    \includegraphics[width=0.95\linewidth]{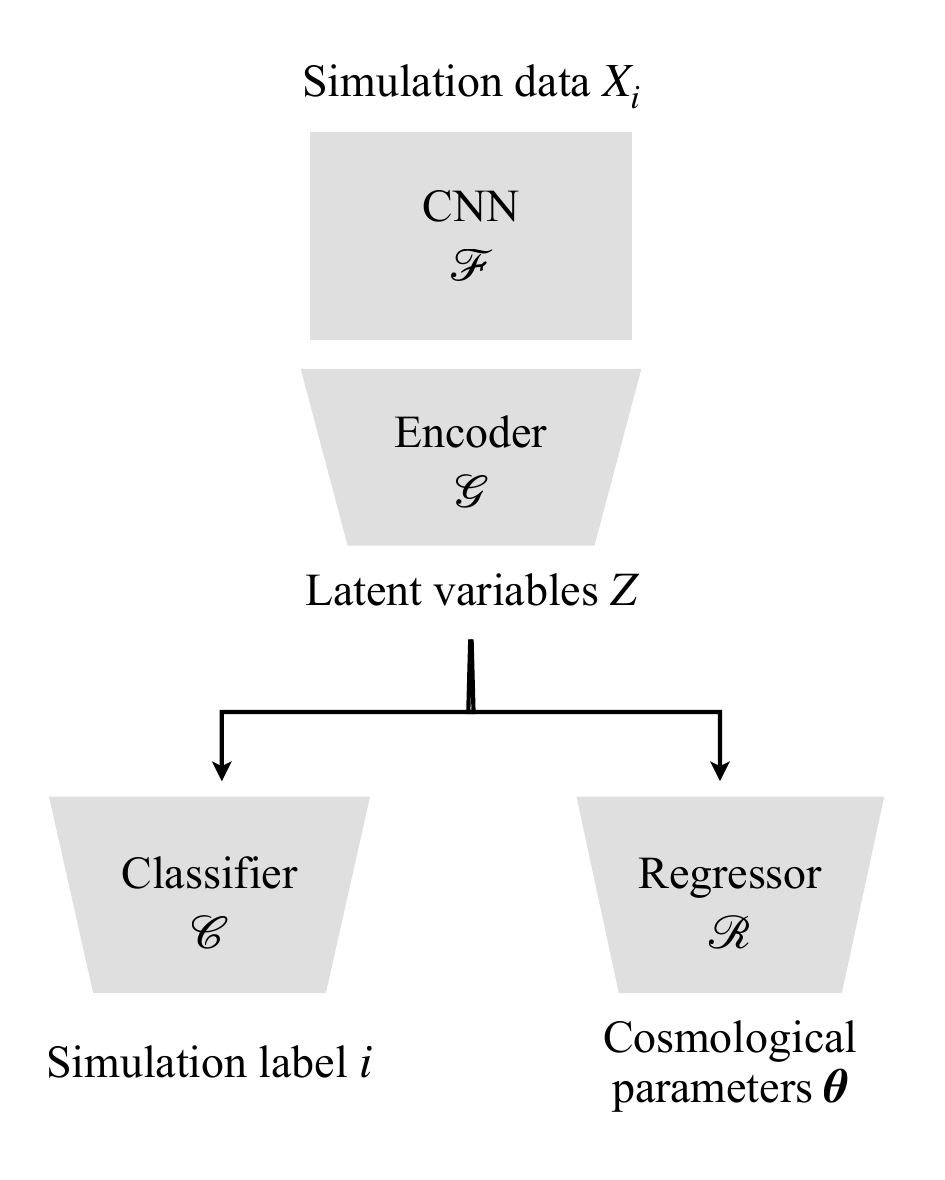}
    \caption{Structure of the neural network. 
    The neural network is comprised of a Convolution Neural Network (CNN) $\mathcal{F}$, an encoder $\mathcal{G}$, a classifier $\mathcal{C}$, and a regressor $\mathcal{R}$.}
    \label{fig:diagram}
\end{figure}

\subsection{Neural Network: \miest{}}
\label{sec:method_neural_network}

The objective of this study is to eliminate information that distinguishes between simulation models, while simultaneously extracting information pertinent to the cosmological parameters.
To achieve this, we utilize methods of compression and de-classification by combining the approaches of deep variational information bottleneck \citep{alemi2017deep} and generative adversarial network \citep{ganNIPS2014_5ca3e9b1}. 
Here, de-classification refers to the procedure in which the machine imposes penalties on the classification mechanism to inhibit the identification of simulation models. 
Hereafter, our machine is referred as to \miest{} (Model-Insensitive ESTimator).

Fig. \ref{fig:diagram} illustrates the structure of the \miest{}. 
The \miest{} is composed of four parts:
\begin{enumerate}
\item The CNN $\mathcal{F}$ takes 2D maps as input and manipulate them into 1D arrays;
\item The encoder $\mathcal{G}$ takes 1D arrays from the CNN and processes (usually compress) their information to latent variables $Z$;
\item The regressor $\mathcal{R}$ outputs the mean and standard deviation of cosmological parameters taking $Z$ as input;
\item The classifier $\mathcal{C}$ predicts labels of simulation models, $i$, and this information is used to train the CNN and the Encoder to omit the classifiable information.
\end{enumerate}

Algorithm \ref{alg:miest} elucidates the comprehensive training procedure. 
The training process is comprised of two distinct phases. 
First, we train the CNN, the encoder, and the regressor using the objective function $L_1$ (refer to Eq.~\ref{eq:loss_function_1} in Sec. \ref{sec:method_neural_network_equations}). 
Secondly, only the classifier is updated utilizing the cross entropy loss $H$.
The rationale for the separate training phases is motivated by our objective to leverage the classifier for the de-classification or removal of classifiable information within the CNN and predominantly the decoder, rather than optimizing for classifier performance.
To this end, it is imperative that we possess a well-trained classifier readily available for the de-classification process. 
By repeatedly applying this procedure over multiple iterations with the same training data, the encoder is trained to generate more compact and compressed latent variables such that they are un-classifiable across different simulations, simultaneously enhancing the accuracy of the regressor.

\RestyleAlgo{ruled}
\SetKwInput{KwInput}{Input}                
\SetKwInput{KwNets}{Neural Network}                
\SetKwInput{KwOutput}{Output}              
\SetKwInput{KwVariable}{Variable}              
\begin{algorithm}[!t]
\label{alg:miest}
\caption{Model Insensitive Estimation}
\KwInput{Training set $\bs{X}$, iterations $N$}
\KwNets{CNN $\mathcal{F}(\bs{X})$, Encoder $\mathcal{G}(Y)$, Classifier $\mathcal{C}(Z)$, $\mathcal{R}(Z)$}
\KwVariable{latent variable $Z$, simulation label $\bs{i}$}
\KwOutput{Mean of cosmological parameters $\bar{\theta}$, Standard deviation of cosmological parameters $\sigma(\theta)$}

\For{$n = 1$ to $N$}{
{\bf Phase I: Regression}\\
Process $\bs{X}$ through the CNN: $Y = \mathcal{F}(\bs{X})$\\
Compress $Y$ through the \tr{encoder}: $Z = \mathcal{G}(Y)$\\
Predict cosmological parameters $(\theta, \sigma) = \mathcal{R}(Z)$\\
Optimize $\mathcal{F}$, $\mathcal{G}$, and $\mathcal{R}$ using gradient decent with $\mathcal{L}_1=J_1 + J_2 - \beta H + \gamma I$\\
{\bf Phase II: Classifier For De-classification}\\
Process $\bs{X}$ through the CNN: $Y = \mathcal{F}(\bs{X})$\\
Compress $Y$ through the encoder: $Z = \mathcal{G}(Y)$\\
Predict simulation labels $\bs{i} = \mathcal{C}(Z)$\\ 
Optimize $\mathcal{C}$ using gradient decent with $\mathcal{L}_2=H$}
\end{algorithm}

\subsubsection{Equations}
\label{sec:method_neural_network_equations}
The objective function $\mathcal{L}_1$ in Algorithm \ref{alg:miest} is written as
\begin{equation}
\label{eq:loss_function_1}
\mathcal{L}_1 = J_1 + J_2 -\beta H +\gamma I,
\end{equation}
where
\begin{align}
\label{eq:loss_function_detail}
    J_1 =& \int_{}^{}\left(\boldsymbol{\theta}-\mathcal{R}_\theta(z)\right)^2p(\theta,z)\,\mathrm{d}\theta\mathrm{d}z,\\
    J_2 =& \int_{}^{}\left(\left(\boldsymbol{\theta}-\mathcal{R}_\sigma(z)\right)^2-\hat{\sigma}(z)^2\right)^2 p(\theta,z)\,\mathrm{d}\theta\mathrm{d}z,\\
    H =&-\int p(z,i)\log \mathcal{C}(z,i)\,\mathrm{d}z\dd i,\\
    I =& I(X;Z) = D_\mathrm{KL}\left(P_{X,Z}||P_{X}\otimes P_{Z}\right).
\end{align}
Here, the latent variables are given through the CNN and the encoder as $Z = \mathcal{G}(\mathcal{F}(\bs{X}))$, and the regressor outputs two values---the mean and the standard deviation---as $\mathcal{R}(z)=\left(\mathcal{R}_{\theta}(z), \mathcal{R}_{\sigma}(z)\right)$.

\tr{The brief descriptions of each loss term are as follows: $J_1$ and $J_2$ are the mean square errors of the mean and variance of the the parameters, that is, for the estimation of the mean and variance of the parameters \citep{jeffrey2020arXiv201105991J}; The cross entropy loss $H$ is used for de-classification to discard the distinguishable information in latent variables as in Fig.~\ref{fig:diagram}; Lastly, the mutual information $I$ is implemented for compression of input data into latent variable motivated by the deep variational information bottleneck \citep{alemi2017deep}, where the information bottleneck method is a framework in information theory used to extract the most relevant information about a target variable from an input data while compressing input data as much as possible.}

\subsubsection{Hyperparmeter and Optimization}
We use {\tt Optuna} for tuning the hyper-parameters within the neural network and the ones used in training such as the structure of the neural networks, dimension of latent variable, dropout, learning rate, and weight decay \citep{optuna10.1145/3292500.3330701}.
{\tt Optuna} is an open-source hyperparameter optimization framework designed to enhance machine learning models by automating the search for optimal hyperparameters. 
It uses a define-by-run paradigm, allowing dynamic search space specification during execution, accommodating complex configurations. 
{\tt Optuna} primarily employs Bayesian optimization, which we adopt in this work, but also offers grid and random search strategies. 
Additionally, it uses pruning techniques to stop unpromising trials early, saving computational resources, and offers visualization tools for optimization analysis. 

The objective in the optimization with {\tt Optuna} is to minimize the regression error $J_1$ and minimize the normalized AUC score\footnote{\label{foot:auc_score}The Area Under the Curve (AUC) score measures binary classification performance by representing the area under the ROC curve, which plots the True Positive Rate against the False Positive Rate at various thresholds. The AUC score ranges from 0 to 1, where 1 indicates perfect performance, 0.5 represents random guessing, and values below 0.5 suggest worse than random performance.} for the de-classification on the validation data.
The normalized AUC score is defined as $\left|\mathrm{AUC\ score} - 0.5\right|$, of whose values range from 0 to 0.5, where 0 indicates non-classifiability and 0.5 represents perfect classification performance. 
Thus, the optimization process is designed to enhance the estimation accuracy of cosmological parameters while inhibiting the classifier's ability to differentiate among the simulation models.

\tr{The free parameters subject to optimization consist of the number of hidden layers, dropout, weight decay, learning rate, and the dimensionality of the latent space. Specifically, the dimensionality of the latent space, as defined by dim($Z$), is sampled from a uniform probability density ranging from 50 to 2000. This range has been chosen based on empirical evidence. 
In this study, the machines trained with and without robustness exhibit a difference in the dimensionality of the latent space, specifically 1375 and 793 dimensions, respectively. This variation in dimensionality may be associated with the quantity of physical information within the input-output pairs, depending on constraints such as the degree of robustness. However, we will not explore this relationship further in the work, as it lies beyond the scope of our current investigation.}

In this work, we choose the one that is most accurate on the test set of the simulations, on which the machine is trained, as our best model.
Presented in the result sections---Sec. \ref{sec:latent_space} and Sec. \ref{sec:performance}---are from the best model.
Further investigation on variance will be discussed in Sec. \ref{sec:convergence_stability}

\subsection{Dimensionality Reduction} 
\label{sec:method_dimension_reduction}
Owing to the inherent high-dimensionality of the latent space, discerning its implications presents a significant challenge.
Accordingly, we employ dimensionality reduction techniques to facilitate a more detailed analysis and visualization of the latent space.

\subsubsection{Uniform Manifold Approximation and Projection}
\label{sec:method_umap}
Uniform Manifold Approximation and Projection (UMAP) is a nonlinear dimensionality reduction technique based on manifold learning and topological data analysis \citep{mapmcinnes2020umapuniformmanifoldapproximation}. 
The goal of UMAP is to reduce the dimensionality of data while preserving as much of the local and global structure as possible. 
The algorithm first constructs a weighted k-nearest neighbor (k-NN) graph, where distances between points in the high-dimensional space are converted into a fuzzy topological representation using a Riemannian metric. 
Once this high-dimensional graph is constructed, UMAP proceeds to optimize it into a lower-dimensional space by minimizing a cross-entropy between the two topological representations, thereby ensuring that the relationships and distances from the original space are preserved in the low-dimensional embedding.

UMAP consists of several parameters, including `n neighbors`, which controls the size of the local neighborhood used for manifold approximation. 
Smaller values of `n neighbors` focus on preserving local structure, whereas larger values help capture more global structures. 
Another important parameter is `min dist`, which determines how closely points are packed together in the low-dimensional space, effectively balancing the preservation of local versus global structures. 
Additionally, UMAP supports various distance metrics, such as Euclidean and cosine, for computing distances between points in the input space.
The selection of these parameters varies between different figures, in alignment with the specific objectives of our analysis. 
The parameters are mentioned in the accompanying text.

\subsubsection{Kernel Principal Component Analysis}
\label{sec:method_kernelpca}
Kernel Principal Component Analysis (KernelPCA) is an extension of traditional Principal Component Analysis (PCA) that allows the extraction of principal components in a higher-dimensional feature space \citep{pca10.5555/646257.685385}. 
KernelPCA employs the kernel trick to map the data into a higher-dimensional space using a kernel function, such as the Radial Basis Function or polynomial kernel. Importantly, this mapping is done implicitly, with the kernel trick computing the dot product between data points in the transformed space without explicitly performing the transformation.

After the kernel function generates the kernel matrix, representing pairwise similarities between data points in the transformed space, the matrix is centered by subtracting the mean to ensure that the data has zero mean in the new feature space. 
PCA is then applied to this centered kernel matrix, where the eigenvectors corresponding to the largest eigenvalues are selected as the principal components. 
These components capture the most significant variance in the data after transformation through a kernel.

In general, UMAP is preferred for its scalability, speed, and ability to preserve local and global structure, making it ideal for visualization and exploratory analysis of large, high-dimensional datasets.
On the other hand, kernelPCA excels in cases where nonlinear data structures are crucial and specific kernel choices can exploit known relationships, but it is limited by scalability issues and requires careful tuning of kernel parameters.
Hence, we utilize UMAP for broad applications due to its versatility, while implementing kernelPCA with the linear kernel specifically to interrogate the behavior of cosmological parameters within the latent space, where established relationships are presumed.

\section{Latent Space}
\label{sec:latent_space}
The aim of this work is to build a robust machine that performs consistently on different models.
However, before diving directly into machine performance, we first investigate the anatomy of \miest{} and then investigate how robustness in \miest{} works.
To this end, we investigate the configuration of the latent space that encapsulates the summary statistics derived from simulations.

The latent space refers to an abstract representation of data that can be obtained as output from the encoder that compresses the simulation maps through CNN (see Sec. \ref{sec:method_neural_network}).
By construction, \miest{} is designed to process the information from the simulations into latent variables so that the latent variables are simultaneously expressive and robust.
Hence, understanding the behaviors and characteristics of the latent space is crucial to analyze the mechanism or performance of \miest{} and ultimately to grasp the intricate relationship between different simulations.

However, the dimension of the latent space ranges from $\mathcal{O}(10^2)$ to $\mathcal{O}(10^3)$. 
Substantial dimensionality poses considerable challenges for latent-space analysis. 
To address this, we employ dimensionality reduction techniques that yield insightful assessments of high-dimensional data (refer to Sec.~\ref{sec:method_dimension_reduction}). 
\tr{The reduced latent space is denoted as $\tilde{\mathcal{Z}}_\mathrm{UMAP}$ and $\tilde{\mathcal{Z}}_\mathrm{kPCA}$, respectively.}
Throughout this paper, we choose three dimensions to which the latent space is reduced. 
First of all, there are at least two axes reserved for $\om$ and $\sig$; it is not necessary that the two $\om$ and $\sig$ axes be orthogonal, but their combinations will be.
Theoretically, there can be a third axis that represents simulation models or classifications of simulations. 
However, it is uncertain whether the classification axis should exist and whether the axis can be an orthogonal basis that does not overlap with other axes.

After examinations with multiple different dimensions for reduction with UMAP and KernelPCA, we have found that dimensions exceeding three cannot make any significant differences \tr{in visualization}; the extra dimensions are redundant with respect to $\om$, $\sig$, and classification. 
Additionally, no axis forms an orthogonal basis for classification, although it consists of a signature of classifications; presumably, it is overlapped with other information.
Hence, we empirically select the three dimensions.
Notice that these results are from UMAP and KernelPCA, implying that there may theoretically still exist an orthogonal basis for classification. 


\begin{figure*}[ht!]
    \centering
    \includegraphics[width=0.98\linewidth]{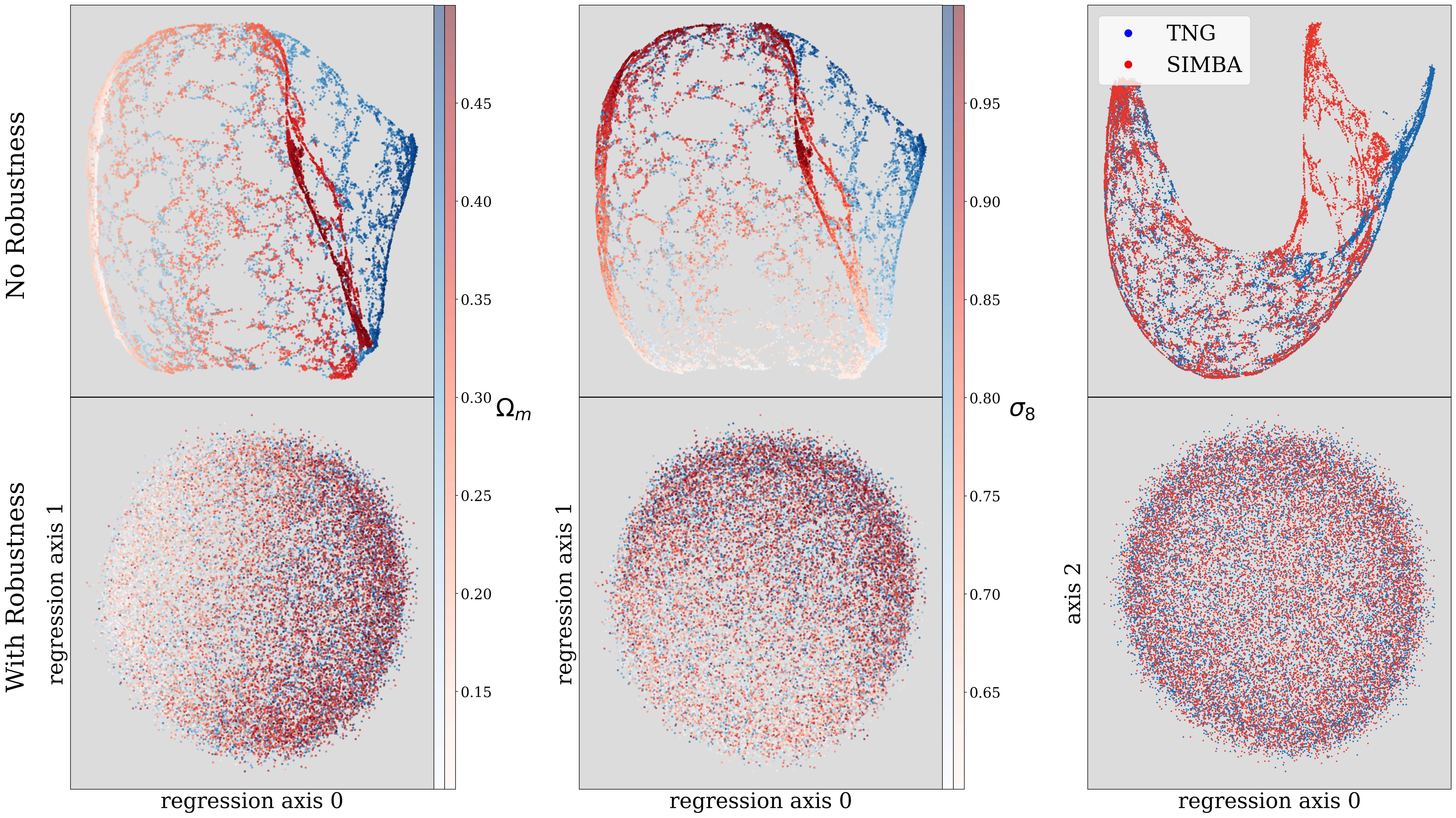}
    \caption{
    Two-dimensional projections of the three dimensional-reduced latent variable \tr{$\tilde{Z}_\mathrm{UMAP}$} using UMAP.
    Since UMAP naturally reduces the dimension of the data, the axes above are totally arbitrary given by UMAP.
    However, one can deduce that due to the nature of the data, there must be at least two axes that represent $\om$ and $\sig$, respectively.
    We denote them as regression axes 0 and 1.
    Each data point is color-coded by their true parameter values $\om$ ({\it first column}) and $\sig$ ({\it second column}).
    This shows that each regression axis is a combination of $\om$ and $\sig$.
    Arguably, we assume that the third axis should be the classification axis that represents the simulation model of the data.
    Here, the reason they are sparse or porous is that we use a test set. 
    Although the `No Robustness' case looks like \tng\ and \simba\ are blended together, it is the structure that two sheets of \tng\ and \simba\ are spanned really close to each other almost in parallel. 
    One can confirm a more clear impact of robustness with the three-dimensional construction of the latent variable space.
    In addition, it is manifested in the zoom-in plot (see Fig.~\ref{fig:latent_tng_simba_zoom}).
    For 3D visualization using the entire dataset, check out the movies of \href{https://youtu.be/_6hqrcqTQ_c}{`No Robustness'} and \href{https://youtu.be/Xd5SIorPA1s}{`With Robustness'}.}
    \label{fig:latent_tng_simba}
\end{figure*}

\begin{figure*}[ht!]
    \centering
    \includegraphics[width=0.98\linewidth]{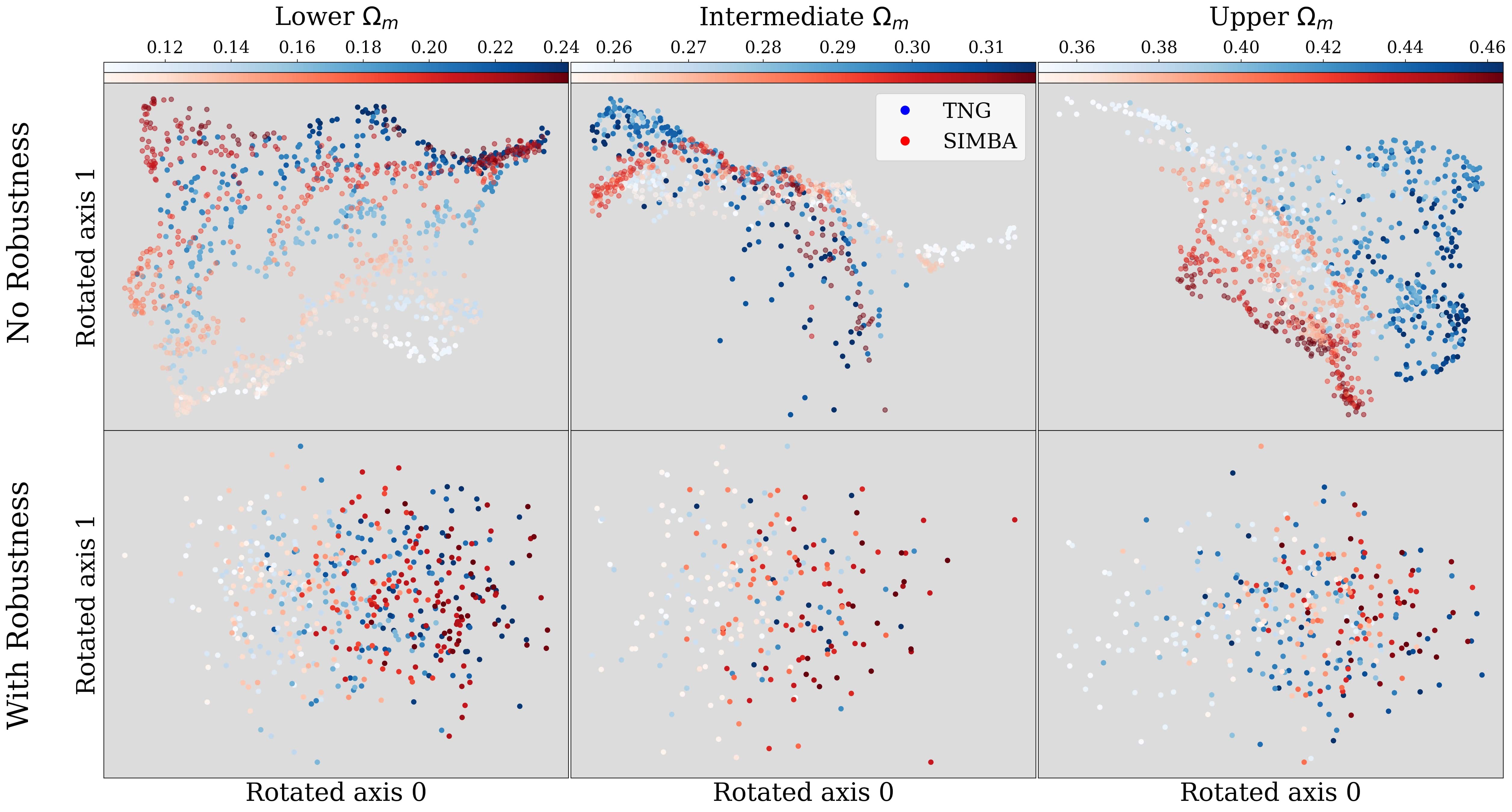}
    \caption{
    Two-dimensional projections of the three dimensional-reduced latent variable space of the test set of \tng\ ({\it blue}) and \simba\ ({\it red}) via UMAP.
    The axes are rotated arbitrarily for better visualization. 
    Each column shows latent variables from three ranges of $\om$: lower, intermediate, and upper, with data points color-coded by their own ranges.
    The {\it upper} panels represent the `No Robustness' case, where \tng\ and \simba\ are classifiable. 
    The {\it lower} panels show the `With Robustness' case, where classification is less evident.}
    \label{fig:latent_tng_simba_zoom}
\end{figure*}

\subsection{How \miest\ Achieves Robustness}
In this section, we use \miest{} trained on \tng{} and \simba{} with and without robustness to investigate how robustness works within the framework of \miest.
Shown in Fig.~\ref{fig:latent_tng_simba} are two-dimensional projections of the three-dimensional latent space \tr{$\tilde{\mathcal{Z}}_\mathrm{UMAP}$} of the whole dataset\footnote{This includes both the training set and test set.} that is dimension reduced through UMAP. 
For UMAP, we use the hyperparameters of 20 neighbors, 0.1 minimum distance, and the `euclidean' metric\footnote{From the experiments, the `canberra' metric is better at capturing clustering. However, the axes for the cosmological parameters become interwoven. For better visualization, we choose `euclidean' for this overview plot.}.
The {\it upper} panels exhibit latent variables \tr{$\tilde{Z}_\mathrm{UMAP}$} of \miest{} trained {\it without} robustness, while the {\it lower} panels show {\it with} robustness.
The machines {\it with} and {\it without} robustness are trained, respectively, and end up having different dimensions of the latent space.
Hence, we conduct the dimensionality reduction separately for each machine.
In other words, there is no direct connection between the upper and lower panels, except that we use the same data set to train the machine.  

The projections in the {\it first column} and {\it second column} are color coded with the true values $\om$ and $\sig$, respectively, where darker colors indicate higher values of the parameters. 
With this in mind, \tng{} ({\it blue}) and \simba{} ({\it red}) have similar trends in terms of cosmological parameters in the latent space along the axes, regardless of robustness.
This suggests that the two axes---denoted as regression axes 0 and 1---are correlated with $\om$ and $\sig$, respectively. 
One must notice that the axes from the dimensionality reduction are completely arbitrary (that is, we have not provided any information on cosmological parameters).
We empirically choose to use the three dimensions as the final reduced dimension.
We anticipate that due to the intrinsic mutual information between the inputs and outputs, there must be two axes reserved for the cosmological parameters $\om$ and $\sig$.
Precisely, there should exist two axes, corresponding to $\om$ and $\sig$, which can be constructed as a combination of the basis vectors within the reduced latent space \tr{$\tilde{\mathcal{Z}}_\mathrm{UMAP}$}. 
For this reason, we denote two axes that are correlated with cosmological parameters as `regression axes 0 and 1' and the other axis as `axis 2'.

The structural characteristics of `No Robustness' are analogous to networks of blood vessels or large-scale structures, being composed of sheets and clumps.
\tng\ and \simba\ represent two closely juxtaposed thin sheets.
Significant separation is observed at elevated values of $\om$ and $\sig$.
Upon meticulous examination, slight misalignments in their clumpy structures become evident, indicating classificationability.
We also investigate the classificationability of latent variables \tr{$Z$} derived from two distinct simulations utilizing the AUC (Area Under the Curve) score.
An AUC score of 0.97 is obtained, demonstrating that the latent variables \tr{$Z$} of the two simulations are clearly classifiable.

Conversely, 'With Robustness' exhibits smooth oval or spherical distributions that are consistent across both the \tng\ and \simba\ datasets, despite the UMAP hyperparameters remaining identical to those in the 'No Robustness', which generally influence the structural features of the reduced space \tr{$\tilde{\mathcal{Z}}_\mathrm{UMAP}$}. 
The distinction between the two simulations becomes visually indistinguishable \tr{in the reduced space $\tilde{\mathcal{Z}}_\mathrm{UMAP}$}. 
Achieving an AUC score\footref{foot:auc_score} of 0.53, it successfully mitigates the classification of simulations, marking a progression towards robustness \tr{also in the full latent space $\mathcal{Z}$}. 
However, it should be noted that the correlation between cosmological parameters and the regression axes weakens compared to the 'No Robustness' case overall. 
This potentially signifies a loss of cosmological information within the latent variables $Z$, as discussed in Sec.~\ref{sec:performance}. 
An important caveat is that the observed clumpiness or any characteristic structures in the reduced latent space \tr{$\tilde{\mathcal{Z}}_\mathrm{UMAP}$} are likely contingent upon the choice of dimensionality reduction methodology, and hence are not physically meaningful.

We turn to the effect of robustness on the test set that the \miest{} has never seen. 
Fig.~\ref{fig:latent_tng_simba_zoom} shows the latent variables \tr{$\tilde{Z}_\mathrm{UMAP}$} divided into three parts with respect to the values of $\om=[0.1,0.25), [0.25, 0.35), [0.35, 0.5]$---denoted lower, intermediate, and upper, respectively.
Each column shows colored data points in latent space \tr{$\tilde{Z}_\mathrm{UMAP}$} for each $\om$ range.
The latent space \tr{$\tilde{\mathcal{Z}}_\mathrm{UMAP}$} is again reduced by UMAP using hyperparameters of 20 neighbors, minimum distance of 0.5, and the `canberra' metric.
The selection of the `canberra' metric accentuates the clustering of simulations more effectively. 
The axes presented have been rotated in an arbitrary manner to improve the visualization of the clustering between \tng\ and \simba\ datasets.

`No Robustness' ({\it upper} panels) shows how \tng\ and \simba\ are configured in latent space \tr{$\tilde{Z}_\mathrm{UMAP}$}.
The most noticeable separation between \tng\ and \simba\ is observed in the upper $\om$ ({\it right} panel), as also seen in Fig.~\ref{fig:latent_tng_simba}. 
The \tng\ ({\it blue}) and \simba\ ({\it red}) are proximal but distinctly segregated on a global scale. 
Meanwhile, clustering is also evident in the lower and intermediate $\om$, though it occurs on a more localized level, with some instances of intermixing noted. 
The AUC scores are 0.93 (lower), 0.98 (intermediate) and 0.99 (upper), indicating the distinctive features of the HI maps for \tng\ and \simba\ manifest most in the higher $\om$. 
In contrast, the "With Robustness" panels ({\it lower}) exhibit no evident clustering or segregation, evidenced by AUC scores of 0.50 (lower), 0.49 (intermediate) and 0.53 (upper). 
Therefore, robustness implementation effectively integrates simulations.

\begin{figure*}[ht!]
    \centering
    \includegraphics[width=0.98\linewidth]{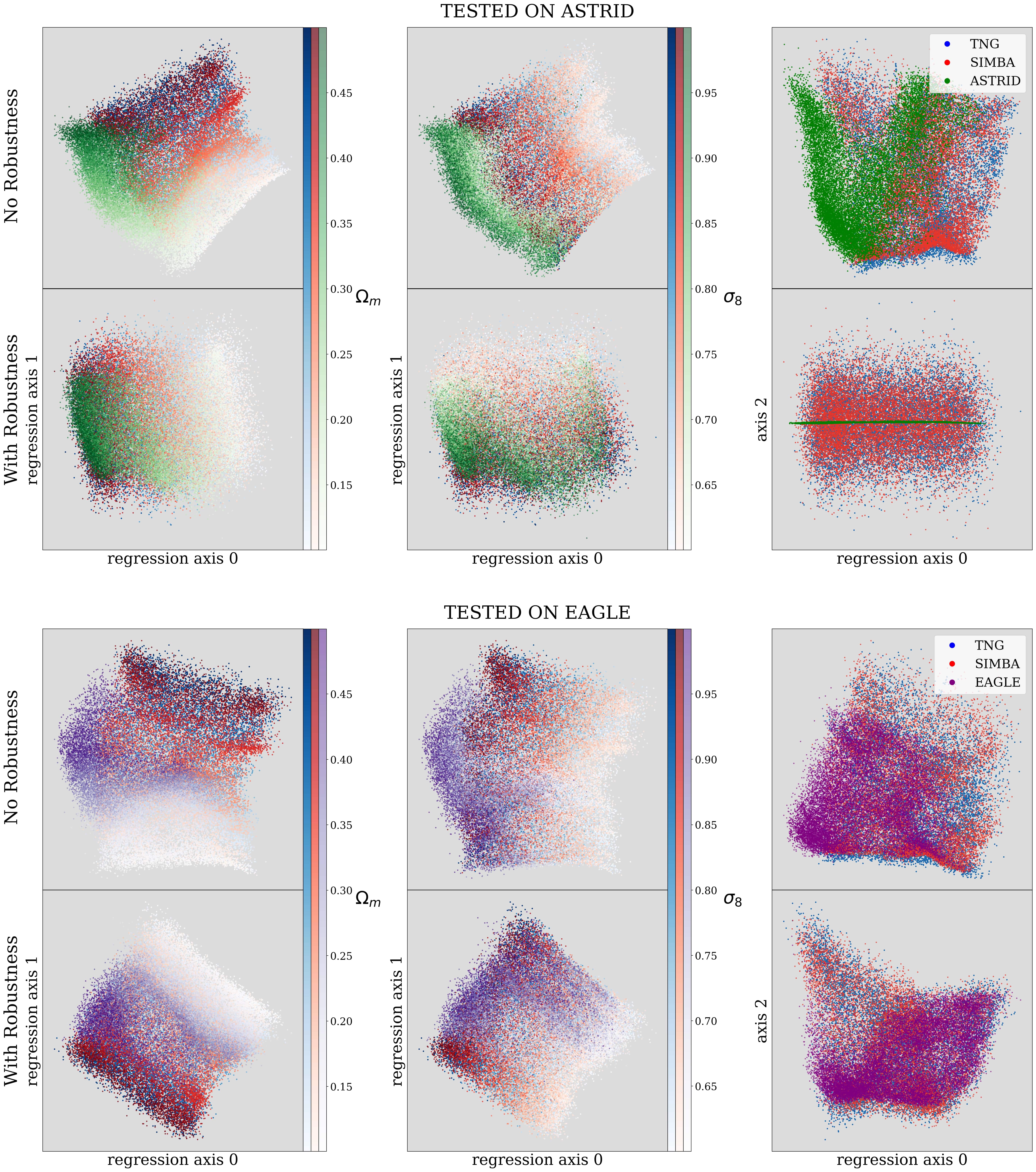}
    \caption{
    Two-dimensional projections of the three dimensional-reduced latent space \tr{$\tilde{\mathcal{Z}}_\mathrm{kPCA}$} using KernelPCA with the linear kernel.
    The \miest\ machine is trained on \tng\ ({\it blue}) and \simba\ ({\it red}) and tested on \astrid\ ({\it green}) and \eagle\ ({\it purple}).
    The presented latent space are constructed from the whole data of \tng\ and \simba\ including the training set and the test set, whereas the latent variables of \astrid\ and \eagle\ has not been seen by the \miest\ machine.
    Each data point is color-coded by its true parameter value of $\om$ ({\it first column}) or $\sig$ ({\it second column}).
    Due to robustness, the latent variables of \astrid{} and \eagle{} are spanned more evenly in the `With Robustness' case ({\it lower} panels) than in `No Robustness' ({\it upper} panels).}
    \label{fig:latent_tng_simba_on_astrid_eagle}
\end{figure*}

\subsection{Effect of Robustness In Unseen Simulations}
We focus our investigation on the influence of robustness on data that the machine has not encountered previously. 
We introduce the unseen datasets, \astrid{} and \eagle{}, into the same machine utilized in the previous section, which has been trained on \tng{} and \simba{}. 
Fig.~\ref{fig:latent_tng_simba_on_astrid_eagle} illustrates the configurations of the latent space \tr{$\tilde{\mathcal{Z}}_\mathrm{kPCA}$} for \astrid\ ({\it top}) and \eagle\ ({\it bottom}), superimposed on the latent variables of \tng\ and \simba.
In contrast to the previous section, kernelPCA with the linear kernel is employed instead of UMAP, often yielding more interpretable results wherein the axes exhibit greater correlation with cosmological parameters (refer to Sec.~\ref{sec:method_kernelpca}). 
The objective of this section is to scrutinize how \astrid\ and \eagle\ are arranged within the latent space \tr{$\tilde{Z}_\mathrm{kPCA}$} spanned by \miest\ trained on \tng\ and \simba, with a particular emphasis on robustness, especially in the context of cosmological parameters of interest for regression. 
Despite minor intricacies that are challenging to discern visually, the variations in \astrid\ and \eagle{}, in response to robustness, exhibit a similar pattern.

Without robustness ({\it upper} panels in each plot), \astrid{} ({\it green}) and \eagle{} ({\it purple}) are placed at the edge of the sheets that \tng\ ({\it blue}) and \simba\ ({\it red}) spans.
The coverage of the unseen data (\astrid{} and \eagle{}) over the training data (\tng{} and \simba{}) is much less than 50\% on average.
This can introduce bias in the regression results since the locus of the latent space \tr{$\tilde{\mathcal{Z}}_\mathrm{kPCA}$} can roughly translate into the predicted values. 
However, since what we are looking at is the reduced space, the actual regression occurs in higher dimensions in a more complex manner. 
The third column that is supposedly related to classification of simulations demonstrates the evident distinction between \astrid\ and the training data, whereas \eagle\ seems to share more similarity with the training data, yet it is classifiable.
We will revisit the similarity between simulations in Sec.~\ref{sec:discussion_difference_in_simulations}

\begin{figure*}[]
    \centering
    \includegraphics[width=0.98\linewidth]{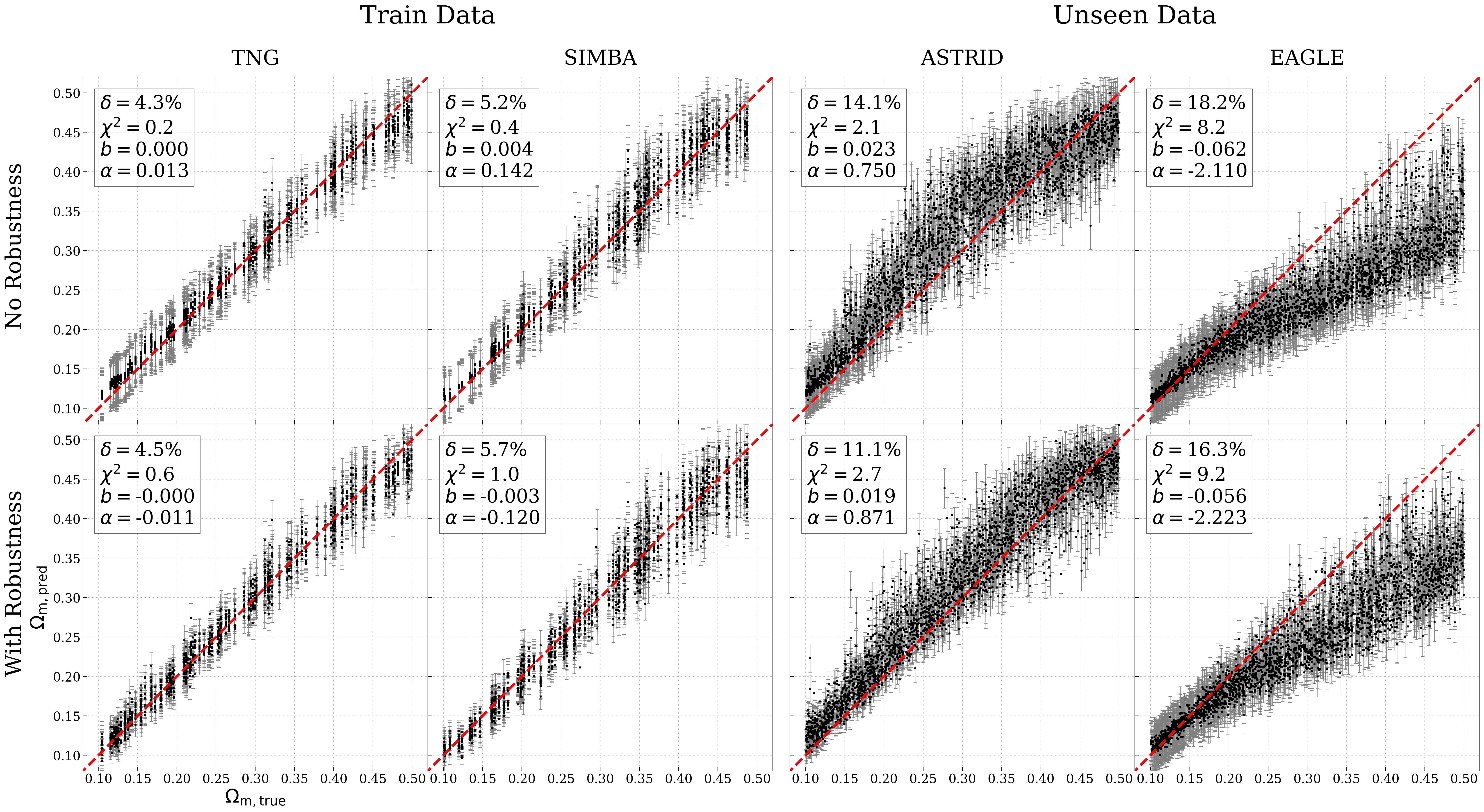}
    \caption{Prediction of $\om$ versus truth {\it with} and {\it without} robustness ({\it lower} and {\it upper}).
    The MIEST is trained on the HI maps of \tng\ ({\it 1st column}) and \simba\ ({\it 2nd column}) and tested on that of \astrid\ ({\it 3rd column}) and \eagle\ ({\it 4th column}).
    Each plot has a relative error $\delta$, $\chi^2$, bias $b$, and $\alpha$.
    Here, $\alpha$ measures how the machine can cover the parameter space with less bias and sufficient confidence intervals (refer to Sec. \ref{sec:performance}).
    }
    \label{fig:one_to_one_om}
\end{figure*}

The incorporation of robustness results in a significant enhancement in coverage for both \astrid\ and \eagle\ ({\it lower} panels in each plot) compared to the case without robustness. 
Consequently, the majority of data points now align with the plane that \tng{} and \simba{} span on the reduced latent space \tr{$\tilde{Z}_\mathrm{kPCA}$}. 
Nonetheless, there remain biases and regions where \astrid{} and \eagle{} barely populate. 
In the case of $\om$ for \eagle{} with robustness ({\it lower left corner}), the distribution of the purple dots in the high $\om$ region is notably sparse, with a discernible bias towards the middle and lower $\om$ values. 
This finding is also consistent with the prediction results depicted in Fig.~\ref{fig:one_to_one_om}, which illustrates the predictions for \eagle{} with robustness (details in Sec. \ref{sec:performance}).
This indicates that these are potential sources of systematic bias and inaccuracies in the prediction.
Overall, the introduction of robustness demonstrates the capacity to enhance predictive accuracy by augmenting the congruence between training simulations and unobserved simulations. 
However, residual biases and potential inaccuracies, which presumably stems from intrinsic properties of simulations, persist to a certain extent.

\begin{figure*}[]
    \centering
    \includegraphics[width=0.98\linewidth]{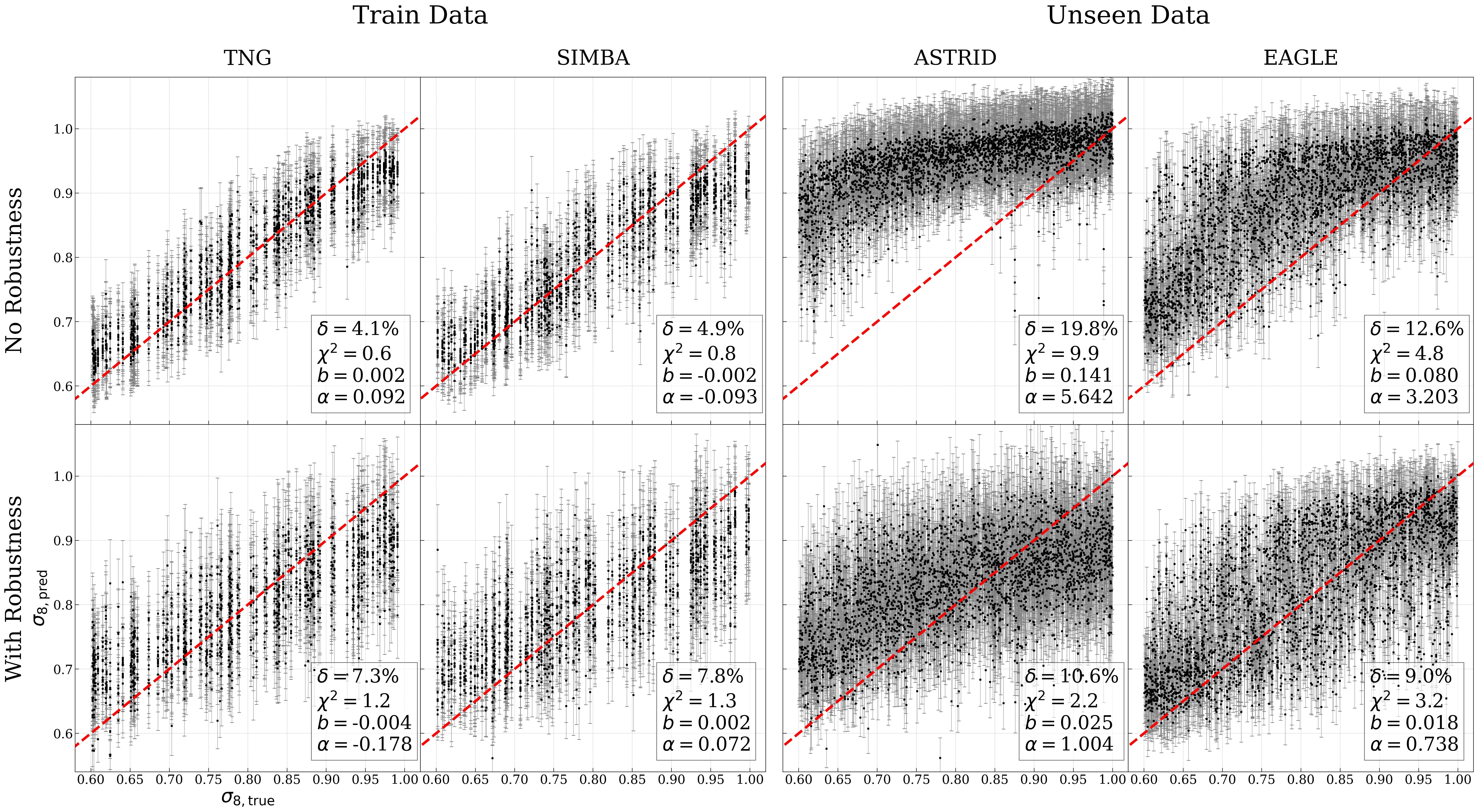}
    \caption{Prediction of $\sig$ versus truth {\it with} and {\it without} robustness ({\it lower} and {\it upper}).
    The MIEST is trained on the HI maps of \tng\ ({\it 1st column}) and \simba\ ({\it 2nd column}) and tested on that of \astrid\ ({\it 3rd column}) and \eagle\ ({\it 4th column}).
    Each plot has a relative error $\delta$, $\chi^2$, bias $b$, and $alpha$.
    Here, $\alpha$ measures how the machine can cover the parameter space with less bias and sufficient confidence intervals (refer to Sec. \ref{sec:performance}).
    }
    \label{fig:one_to_one_sig}
\end{figure*}

\begin{figure*}[ht]
    \includegraphics[width=0.98\linewidth]{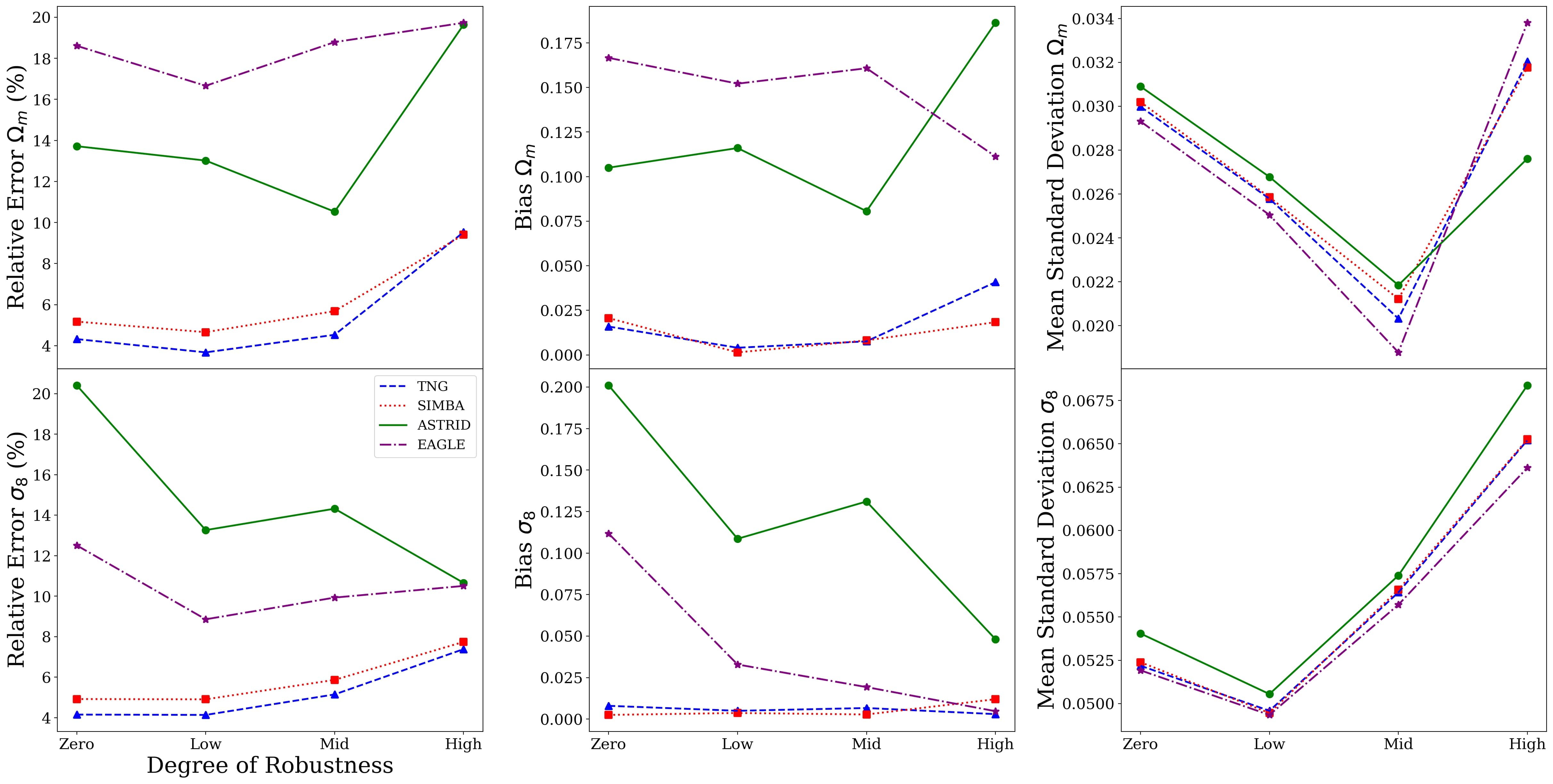}
    \caption{Relative errors of $\om$ ({\it top}) and $\sig$ ({\it bottom}) versus the degree of robustness.
    The MIEST is trained on the HI maps of \tng\ ({\it blue dashed}) and \simba\ ({\it red dotted}) and tested on that of \astrid\ ({\it green solid}) and \eagle\ ({\it purple dot-dashed}).
    }
    \label{fig:degree_of_robustness}
\end{figure*}

\section{Performance}
\label{sec:performance}
In this section, we dive into the robustness performance in an explicit manner. 
Fig.~\ref{fig:one_to_one_om} and \ref{fig:one_to_one_sig} illustrate the predictions of $\om$ and $\sig$, respectively, versus truth with and without robustness using \tng, \simba, \astrid, and \eagle.  
\tng{} ({\it first} column) and \simba{} ({\it second} column), on which the machine is trained, demonstrate the precision of the machine with respect to the robustness.  
We also provide in the figures the values of four metrics that measure the performance of \miest{}: relative error $\delta$, chi-square $\chi^2$, bias $b$, and generalization $\alpha$.
$\chi^2$ is defined by $\left<\theta_\mathrm{true}-\hat\theta_\mathrm{pred}|/\hat\sigma_\mathrm{pred}\right>$, where $\left<\cdot\right>$ and $\hat\sigma_\mathrm{pred}$ are the average over the test set the standard deviation given by \miest{} (refer to Eq. \ref{eq:loss_function_1} and \ref{eq:loss_function_detail}).
The generalization $\alpha$ is defined as $\left<b/\hat\sigma_\mathrm{pred}\right>$.
The generalization $\alpha$ is designed to achieve a minimal value when the confidence interval of the prediction covers a broader parameter space with minimal bias; in other words, the model exhibits less bias and larger standard deviation compared to the bias.
However, this metric does not directly measure the predictive power.

In the absence of robustness, the relative errors of the model for \tng{}(\simba{}) are quantified as 4.3\%(5.3\%) for $\om$ and 4.2\%(4.9\%) for $\sig$. 
The slight difference in accuracy between two simulations originates from the intrinsic amount of information that each simulation carries, as discussed in Sec.~\ref{sec:discussion_difference_in_simulations}.
However, the introduction of robustness results in performance degradation, increasing the relative errors to 4.5\%(5.6\%) for $\om$ and 7.4\%(7.9\%) for $\sig$.
During the implementation of robustness, the machine is compelled to disregard information exclusive to individual simulations, leading to an unavoidable decline in performance.
However, the both values of bias and $\alpha$ decrease with robustness from $4\times10^{-4}$ to $-2\times10^{-4}$ ($4.3\times10^{-3}$ to $-2.7\times10^{-3}$) for bias and $1.5\times10^{-3}$ to $-1.2\times10^{-3}$ ($0.14$ to $-0.13$) for $\alpha$, indicating that the machine yields a more uniform prediction across the parameter space than without robustness.
This suggests that the incorporation of robustness not only has the potential to enhance the generalization for unseen data but also augments the intrinsic robustness for the train data itself.

In the case of the unseen simulations, \astrid\ ({\it third} column) and \eagle\ ({\it fourth} column) show relative errors of 14.1\% and 18.3\% for $\om$ and 20.2\% and 12.9\% for $\sig$, respectively.
The substantial precision discrepancy between the train data and the unseen data can be attributed to the configurations of the latent space \tr{$\tilde{\mathcal{Z}}_\mathrm{kPCA}$} depicted in the `No Robustness' cases of Fig.~\ref{fig:latent_tng_simba_on_astrid_eagle}, which elucidates the evident bias present in \astrid{} and \eagle{} as opposed to \tng{} and \simba{}.
However, by incorporating robustness, the relative errors of \astrid{} and \eagle{} has been improved by 23\% and 11\% for $\om$ and 47\% and 29\% for $\sig$, respectively.
Furthermore, $\chi^2$ is reduced by $2\sim3$ times, with biases decreased by a factor of 7 for \astrid{} and a factor of 4 for \eagle{}, and the values of $\alpha$ are enhanced approximately four-fold for both.
These improvements correlate with the alterations in the latent space \tr{$\tilde{\mathcal{Z}}_\mathrm{kPCA}$} attributed to robustness (see `With Robustness' of Fig.~\ref{fig:latent_tng_simba_on_astrid_eagle}). 
This substantiates the pivotal role of robustness in increasing performance.

In addition to performance enhancement, two noteworthy observations can be made. 
First, the improvement manifests itself with greater prominence in $\sig$ compared to $\om$.
In the absence of robustness, the predictive accuracy of $\sig$ within both \astrid{} and \eagle{} exhibits substantial bias, largely yielding values in the range of 0.8 to 1.0 for $\sig$, irrespective of the actual true value.
However, the implementation of robustness remarkably mitigates such biases.
The predictions have become increasingly aligned with the median values on average, indicating a reduction in bias, and exhibit a higher degree of correlation with the actual true values.
On the other hand, the improvement in $\om$ is relatively minor compared to $\sig$.
Therefore, two conclusions can be drawn: 
a) In the absence of robustness, $\sig$ is substantially biased to their own simulations.  
b) Nevertheless, $\sig$ encompasses a substantial amount of both common and distinguishable information compared to $\om$, since $\om$ shows limited potential for improvement with respect to robustness.

The second noteworthy observation is that the prediction of $\om$ in \eagle{} is significantly biased regardless of the presence of robustness. 
This underscores the inherent divergence in the physical models, as shown in the bottom left panel of Fig.~\ref{fig:latent_tng_simba_on_astrid_eagle}, which illustrates the substantial absence of data points for \eagle{} ({\it purple}) in the high $\om$ region both with and without robustness.
The limitations associated with attaining robustness and divergence of the simulation models will be discussed in greater detail in \ref{sec:discussion_difference_in_simulations}.

We now investigate how performance can change with respect to the degree of robustness.
As discussed in Sec.~\ref{sec:method_neural_network}, the magnitude of de-classification and compression can be modulated by the hyperparemeters $\beta$ and $\gamma$, respectively, in the loss function (see Eq. \ref{eq:loss_function_1})
De-classification and compression together play a significant role in robustness.
Fig.~\ref{fig:degree_of_robustness} shows how the relative error $\delta$ ({\it left}), bias $b$ ({\it middle}), and standard deviation $\sigma_\mathrm{pred}$ ({\it right}) of $\om$ ({\it top}) and $\sig$ ({\it bottom}) changes with respect to the degree of robustness, demonstrating qualitatively the relationship between robustness and hyperparameters.
There are four different levels of robustness on the x axis: zero ($(\beta, \gamma)=(0,0)$), low ($(\beta, \gamma)=(10^{-3},10^{-4})$), mid ($(\beta, \gamma)=(10^{-2},10^{-3})$), and high ($(\beta, \gamma)=(10^{-1},10^{-2})$). 

We can categorize the simulations into those used for training (\tng{} and \simba{}) and those reserved for testing (\astrid{} and \eagle{}). 
For the simulations employed in the training phase, the relationship between relative error and robustness is monotonic. 
An increase in robustness correlates with a reduction in predictive power due to loss of information through compression and de-classification processes.
In principle, such a trend becomes manifest in the pursuit of robustness, as increments in $\gamma$ and $\beta$ result in gradient descent being increasingly influenced by compression $I$ and de-classification $H$, rather than the regression objective function $J_1$ (refer to Sec.~\ref{sec:method_neural_network_equations}). 
This phenomenon is conceptually understandable, as it involves the omission of information that, while distinguishable, may still exhibit correlation with cosmological parameters.

In the case of \astrid{} and \eagle{}, the accuracy in relation to the degree of robustness progressively enhances until reaching a turning point. 
This suggests the existence of one or more optimal points characterized by maximal robustness, necessitating an exhaustive parameter search to identify.
Nevertheless, we have not conducted such exploration due to the complexity and high dimensionality inherent in the hyperparameters, which encompass those pertinent to the optimization of the neural network.
The turnaround occurs relatively earlier for $\om$ than $\sig$, indicating that $\om$ retains less common information or is more sensitive to compression and de-classification processes, thereby resulting in its premature degradation relative to $\sig$.
Furthermore, biases exhibit analogous patterns.
Conversely, the standard deviations demonstrate unexpected outcomes. 
Typically, the loss of information, as evidenced by the relative error in \tng{} and \simba{}, would be associated with an increase in standard deviations. 
However, the observed standard deviations reveal a decline followed by abrupt increases at certain points.

\section{Discussion}
\label{sec:discussion}

\subsection{On Robustness}
\label{sec:discussion_robustness}
The definition of robustness varies depending on the context of the problem. 
This particular study is dedicated to the development of a machine learning model that demonstrates consistent performance across various cosmological simulations, particularly in the context of estimating cosmological parameters from HI maps.
To this end, we have tuned the neural network to remove distinct information but preserve common information that is, importantly, correlated with cosmology.
In this sense, achieving robustness in this work can be defined as finding common information, or latent space, that is both physical and universal to any simulation, with the important assumption that all simulations have physical integrity at a certain level. 
In this context, the latent variables are not required to be directly observable physical quantities, but rather should encompass any physical features of the simulations.

Imagine the feature space $\mathcal{H}_{i}$ of the physical features $h_i$ that can be found in simulation $i$. 
There can be a subset of features $\mathcal{C}_i$ that are correlated with the parameter of interest $\btheta$. That is, $\mathcal{C}_i=\{h_i|I(h_i;\btheta)>0\}$, where the mutual information $I(X;Y)\equiv D_\mathrm{KL}(P_{(X,Y)}\Vert P_X \otimes P_{Y})$ and $D_\mathrm{KL}$ is the Kullback-Leibler divergence.
In the conventional, regular estimator, the objective is to find an optimal subset of features $\mathcal{M}_i \subset \mathcal{C}_i$.
Then, our goal is to find a subset $\mathcal{R} = \{h|h\in \mathcal{A}, \mathcal{A} = \bigcap_{i} \mathcal{C}_i\}$.
Here, the subscript $i$ is no longer needed in $\mathcal{R}$ since the features in $\mathcal{R}$ should be universal, i.e.~in common between all simulations.
In other words, this subset represents the collection of {\it robust} physical features.
We can also define a subset $\mathcal{R}_{ab} = \{h|h\in \mathcal{A}, \mathcal{A} = \bigcap_{i=\{a,b\}} \mathcal{C}_i\}$. 
$\mathcal{R}_{ab}$ represents the subset of the robust features $h$ that are common only in simulations $a$ and $b$.
Note that since the robust subset $\mathcal{R}$ is comprised of $\mathcal{C}_i$ whose input features (in this work, HI map) are correlated with the target parameters (here, cosmological parameters), {\it robustness} is conditioned on given inputs and outputs; e.g.~
our model is only robust on cosmological parameters with respect to HI maps.

With this, we consider the case where we train a model on simulations $[a\ldots m]$ with robustness. 
Ideally, we can have 
 a robust model that has learned $\mathcal{R}_{[a\ldots m]}$ to the fullest.
By definition, the simulations are not distinguishable on $\mathcal{R}_{[a\ldots m]}$. 
In other words, on $\mathcal{R}_{[a\ldots m]}$, the cross entropy $H(P_a, P_b)$ should be the same as the entropy $H(P_a)$ and $H(P_b)$ for any two simulations a and b (i.e., $H(P_a, P_b) = H(P_a)=H(P_b)$).
In practice, however, it is nearly impossible to obtain solely $\mathcal{R}_{[a\ldots m]}$ as whole. 
Rather, during training, it is more likely to get $\mathcal{R}'_{[a\ldots m]} \cup \mathcal{H}'_{[a\ldots m]}$ where $\mathcal{R}'_{[a\ldots m]}\subset \mathcal{R}_{[a\ldots m]}$ and an arbitrary subset $\mathcal{H}'_{[a\ldots m]} = \cup_{i=[a\ldots m]}\mathcal{H}'_i$ for $\mathcal{H}'_i\subset \mathcal{H}_i$ that causes noise. 
Then, the leakage $\mathcal{H}'_{[a\ldots m]}$ eventually lead to classifiability, namely $H(P_a, P_b)>H(P_a)$ since the features $h_i'\in \mathcal{H}_i'$ are distinguishable across different simulations.
This phenomenon is observed in the AUC score trained with robustness.

In addition, we have used relative errors to measure the performance of \miest{}.
For instance, the relative error on \astrid{} is related to the mutual information between the intersection---$\mathcal{R}'_\mathrm{[\tng{}, \simba{}]} \cap \mathcal{R}_\mathrm{[\tng{},\simba{},\astrid{}]}$---and cosmological parameters.
In contrast to the AUC score that accounts for distinguishability, the relative error on the unseen simulations has a theoretical upper bound, which
 is determined by the mutual information between the intersection---$\mathcal{R}_\mathrm{[\tng{},\simba{}]} \cup \mathcal{C}_\mathrm{\astrid{}}$---and cosmological parameters.
The intersection represents the degree of overlap among \tng{}, \simba{}, and \astrid{}, reflecting the inherent properties of simulations irrespective to \miest{}.

Furthermore, during the training phase, the fact that \miest{} lacks information about \astrid{} results in unstable outputs. 
In practice, $\mathcal{R}'$ of the \miest{}---a subset of $\mathcal{R}_\mathrm{[\tng,\simba]}$---is determined in a stochastic manner due to inherent randomness or imperfections in the training processes.
The inherent stochasticity in the training processes introduces randomness into the intersection between $\mathcal{R}'_\mathrm{[\tng{}, \simba{}]}$ and $\mathcal{R}_\mathrm{[\tng{},\simba{},\astrid{}]}$, which in turn determines the performance on \astrid{}. 
This phenomenon arises since \miest{} lacks information about where \astrid{} should lie in $\mathcal{R}_\mathrm{[\tng,\simba]}$.
Consequently, this randomness inevitably results in inherent instability and divergence in the outcomes of \miest{} or robustness studies, which is discussed in detail next.

\begin{figure*}[ht!]
    \centering
    \includegraphics[width=0.98\linewidth]{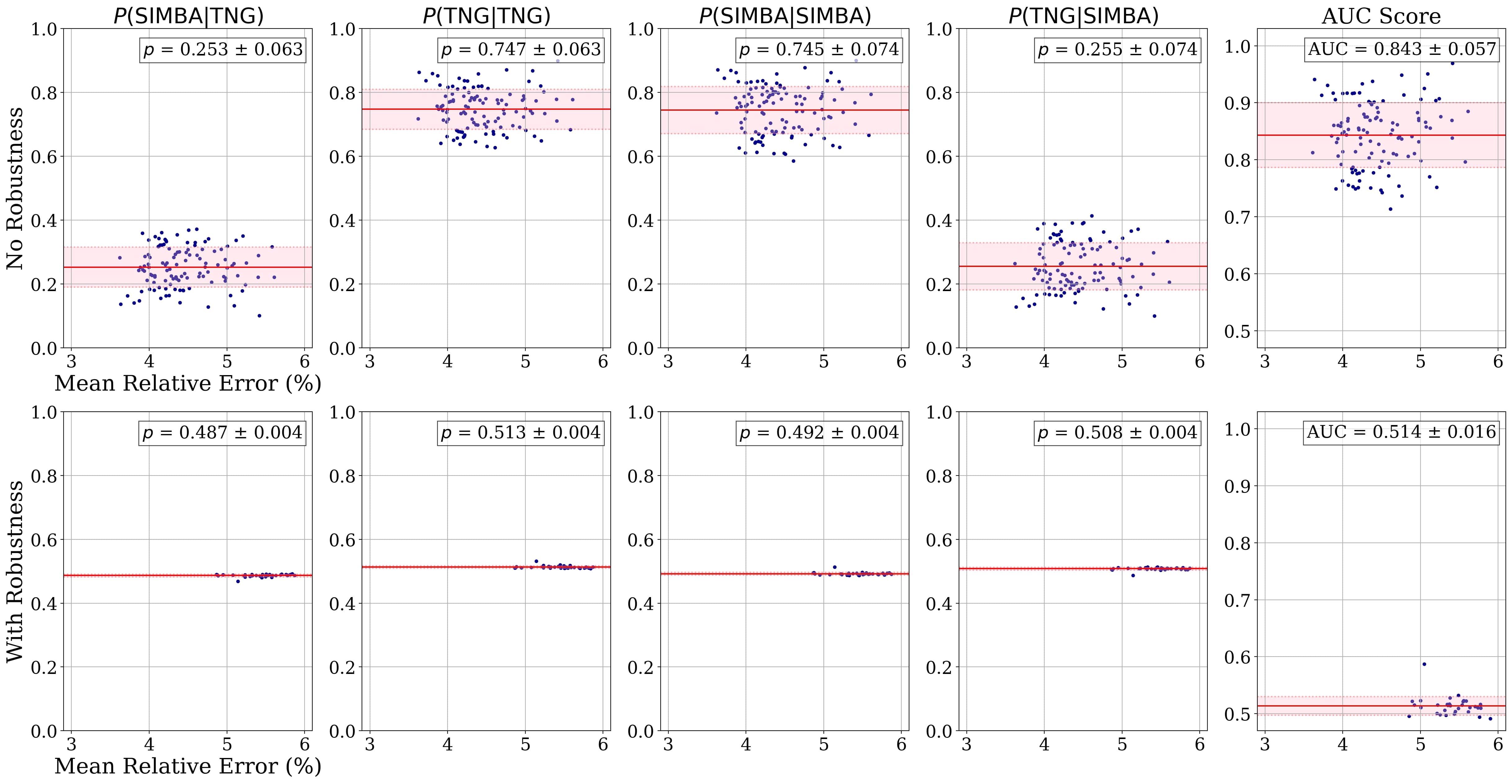}
    \caption{Probability of a simulation model $a$ given latent variables of a certain model--- $Z_b$, $P(a|Z_b)$, and AUC score {\it with} and {\it without} robustness.
    $P(a|Z_b)$ represents the probability of \miest{}, which is trained on \tng{} and \simba, identifying a given input as either \tng{} or \simba{}.
    With $P(\mathrm{\tng}|Z_b)\sim0.5\sim P(\mathrm{\simba}|Z_b)$, robustness ensures that the \miest{} identify the input to be either \tng{} or \simba{} regardless of its origin.
    }
    \label{fig:stability_robustness}
\end{figure*}

\begin{table}[t!]
    \centering
    \caption{Summary of the relative error $\delta$, bias $b$, and generalization $\alpha$ on \astrid{} and \eagle{}. 
    The values presented are averages across multiple \miest{}s trained with and without robustness to account for the effect of randomness in robustness.
    Lower numbers indicate superior performance.
    The superior values among 'No Robustness' and 'With Robustness' are highlighted in bold.}
    \begin{tabular}{|c|c|c|c|c|c|}
        \hline
        \multicolumn{2}{|c|}{} & \multicolumn{2}{c|}{\astrid{}} & \multicolumn{2}{c|}{\eagle{}} \\
        \hline
        \multicolumn{2}{|c|}{Robust.} & Without & With & Without & With\\
        \hline
        \hline
        \multirow{3}{*}{$\om$}&$\delta$ & 15.7$\pm$2.8 & {\bf 14.1$\pm$2.6} & {\bf 16.9$\pm$2.0} & 17.9$\pm$2.4 \\
        \cline{2-6}
        & $10b$   & 1.5$\pm$0.3 & {\bf 1.1$\pm$0.4} & 1.5$\pm$0.3 & {\bf 1.2$\pm$0.5}\\
        \cline{2-6}
        &$\alpha$ & 6.1$\pm$1.6 & {\bf 4.9$\pm$1.8} & 6.7$\pm$1.5 & {\bf 4.75$\pm$2.4} \\
        \hline
        \hline
        \multirow{3}{*}{$\sig$}&$\delta$ & 14.5$\pm$1.9 & {\bf 12.8$\pm$0.5} & 12.7$\pm$1.0 & {\bf 11.8$\pm$1.0}\\
        \cline{2-6}
        & $10b$   & 1.3$\pm$0.3 & {\bf 0.2$\pm$0.1} & 0.9$\pm$0.3 & {\bf 0.2$\pm$0.1}\\
        \cline{2-6}
        &$\alpha$ & 3.1$\pm$0.6 & {\bf 0.2$\pm$0.1} & 2.1$\pm$0.6 & {\bf 0.15$\pm$0.10}\\
        \hline
    \end{tabular}
    \label{tab:stability_robustness}
\end{table}

\subsection{Convergence and Stability of Robustness}
\label{sec:convergence_stability}
In this section, we investigate the convergence as well as the stability of the robustness of our models.
In this context, convergence is described as the phenomenon that the metrics accounting for robustness---e.g., AUC score, performance on \astrid{} and \eagle---progressively approach a stable point as the self-performance of \miest{} on the training simulations improves.
Similarly, stability refers to the variance of these robustness metrics in comparison to those of the machine without robustness. 
For instance, if the variance of the machine with robustness is so large to the extent that it overlaps with the machine without robustness, then it may be considered highly unstable.
Throughout this section, we use \miest{} that is trained on \tng{} and \simba.

Fig.~\ref{fig:stability_robustness} displays the changes in distinguishability of the latent variables {\it with} (bottom) and {\it without} (top) robustness using AUC score (right column) and evidence probabilities (other columns). 
The evidence probabilities in Bayesian statistics refers to the probability of observing the data given all possible values of the parameters within a model.
In this context, a probability of simulation models is defined as the evidence.
We present the outputs of over one hundred \miest{}s, each trained individually and selected based on a mean relative error of $\om$ and $\sig$ below 6\%, represented by the blue dots.
The evidence probability $P(a|Z_b)$ represents the probability of an arbitrary simulation model $a$ given the latent variables of an arbitrary model, $Z_b$.
In other words, this value represents the probability of a given latent variable $Z_b$---generated by the HI maps of simulation $b$---being classified as $a$.
To compute this, we train a random forest classifier using latent variables $Z$ and simulation labels $a$.
The random forest subsequently estimates the probability of simulation models as a function of the latent variables.

`No Robustness' ({\it upper panels}) shows an AUC score of $\sim0.84$ and evidence probabilities of $P(A|A)\sim0.75$ and $P(B|A)\sim0.25$, where $P(A|A)$ and $P(B|A)$ represent the cases where two simulations are the same or are different, respectively.
This indicates that the `\miest' without robustness can classify the simulations, yet there is some overlap of around 25\% in terms of the evidence probability.
The standard deviations of these values are approximately over 10\%, which is not very stable.
For example, some cases demonstrate a probability of 0.4 or 0.6, indicating that \tng{} and \simba{} are not distinguishable to some extent.
Furthermore, there are no evident signs of convergence in the probabilities and the AUC scores towards smaller mean relative errors.
Consequently, training \miest{} without robustness yields unstable and divergent outputs in terms of distinguishability between \tng{} and \simba{}.

Meanwhile, `With Robustness' ({\it lower panels}) shows consistent results across different \miest{}s, demonstrating the AUC score of $\sim0.51$ and $P(A|A)\sim0.51$, and $P(B|A)\sim0.49$. 
This evidently indicates the robustness succeeds in removing the distinct model-specific information so that they are indistinguishable in the latent variables.
Moreover, the standard deviations of the probabilities are less than 1\% while that of the AUC score is $\sim2\%$, which is a huge improvement from `No Robustness'.
This also can evidently be considered convergent.
Overall, the implementation of robustness results in the stable, convergent \miest{}.

We now move on to the performance on the unseen simulations---\astrid{} and \eagle{}.
Tab.~\ref{tab:stability_robustness} summarizes the mean of the relative errors $\delta$, biases $b$, and generalization $\alpha$ and their standard deviations for $\om$ and $\sig$ (refer to the first paragraph of Sec. \ref{sec:performance} for the definitions).
For \astrid{}, it is evident that robustness improves the performance in every metric by more than 10\% in terms of the mean values. 
However, for $\om$ of \astrid{}, the standard deviations are more than 10\%, leading to overlaps between `No Robustness' and `With Robustness'.
This suggests that the improvements in $\om$ exhibit instability, implying that despite certain improvements, the influence of randomness is still significant.
In contrast, the improvements in $\sig$ are more significant and stable. 
Notably, the generalization performance is enhanced by a factor of 15, in line with Fig. \ref{fig:one_to_one_sig}.

The case of \eagle{} displays a similar tendency to \astrid{}, albeit with relatively fewer overall improvements.
A key difference in $\om$ is the inconsistency where the incorporation of robustness results in a higher relative error, deteriorating predictive performance; however, the robustness leads to improvements in the bias and generalization, despite being unstable.
This demonstrates that while robustness in \miest{} results in a loss of predictive power accompanied by increased scatter, it in return enhances generalization, enabling a more uniform coverage of the parameter space with heightened levels of confidence.
This phenomenon is illustrated in Fig~.\ref{fig:latent_tng_simba_on_astrid_eagle}, which demonstrates that robustness enhances the coverage of \eagle{}, albeit at the cost of inducing misalignments.

The convergence and (in-)stability arise from both the inherent characteristics of various simulations and the influence of stochastic elements in the training process.
Ideally, if one can train a machine such that it removes all the distinct features, yet retaining common features, i.e. the latent space of the machine represents exactly $\mathcal{R}_{[\mathrm{\tng},\mathrm{\simba}]}$ in the previous section, this machine is considered convergent and stable, providing consistent results. 
In practice, however, due to stochasticity and incompleteness of the training process, the latent space results in representing only subset of $\mathcal{R}_{[\mathrm{\tng},\mathrm{\simba}]}$, leading to potential issues in convergence and stability.
Furthermore, the performance on \astrid{} (\eagle) is determined by the overlap between the intersection of \tng{} and \simba, $\mathcal{R}_{[\mathrm{\tng},\mathrm{\simba}]}$, and \astrid{} (\eagle{}). 
However, since the training lacks the information about \astrid{} and/or \eagle{}, the extent of the overlap remains completely unaccounted for in the training process, i.e., it is determined stochastically.
Consequently, the performance on unseen simulations tends to be unstable and inherently unable to converge, posing significant hurdles in the robustness studies.

\begin{table*}[]
\centering
\caption{Summary of probability of a simulation model $A$ given latent variables of a certain model--- $Z_B$, $P(A|Z_B)$, and AUC score {\it without} robustness.
$P(A|Z_B)$ represents the probability of \miest{}, which is trained on Simulation A and Simulation B, identifying a given input as either Simulation A or Simulation B.
With $P(\mathrm{B}|Z_A)\sim0.5\sim P(\mathrm{A}|Z_B)$, robustness ensures that the \miest{} identify the input to be either Simulation A or Simulation B regardless of its origin.
This analysis indicates that \tng{}, \astrid{}, and \eagle{} share more common features, whereas \simba{} is an outlier. 
The latent spaces of this result can be found in Fig.~\ref{fig:simulations_full_combinatory}.}
\begin{longtable*}{|c|c||c|c|c|c|c|}
\hline
\makebox[2.7cm]{Simulation A} & \makebox[2.7cm]{Simulation B} & \makebox[2cm]{$P(B|Z_A)$} & \makebox[2cm]{$P(A|Z_A)$} & \makebox[2cm]{$P(B|Z_B)$} &  \makebox[2cm]{$P(A|Z_B)$} & \makebox[2cm]{AUC Score}\\
\hline
\hline
\endfirsthead
\hline
\endhead
\hline
\endfoot
\hline
\endlastfoot
    \tng{}  & \simba  & $0.28\pm0.08$ & $0.72\pm0.08$ & $0.72\pm 0.08$ & $0.28\pm0.09$ & $0.83\pm0.07$\\
    \hline
    \tng{}  & \astrid &  $0.45\pm0.02$ & $0.55\pm0.02$ & $0.51\pm 0.02$ & $0.49\pm0.02$ & $0.58\pm0.04$\\
    \hline
    \tng{}  & \eagle  & $0.45\pm0.04$ & $0.55\pm0.04$ & $0.52\pm 0.04$ & $0.48\pm0.04$ & $0.59\pm0.07$\\
    \hline
    \simba  & \astrid & $0.30\pm0.06$ & $0.70\pm0.06$ & $0.66\pm 0.07$ & $0.34\pm0.07$ & $0.81\pm0.08$\\
    \hline
    \simba  & \eagle  & $0.29\pm0.06$ & $0.71\pm0.06$ & $0.71\pm 0.06$ & $0.29\pm0.06$ & $0.82\pm0.05$\\
    \hline
    \astrid & \eagle  & $0.47\pm0.02$ & $0.53\pm0.02$ & $0.52\pm 0.02$ & $0.48\pm0.02$ & $0.57\pm0.04$\\
\end{longtable*}
\label{tab:two_simulations} 
\end{table*}

\begin{figure*}
    \includegraphics[width=0.98\linewidth]{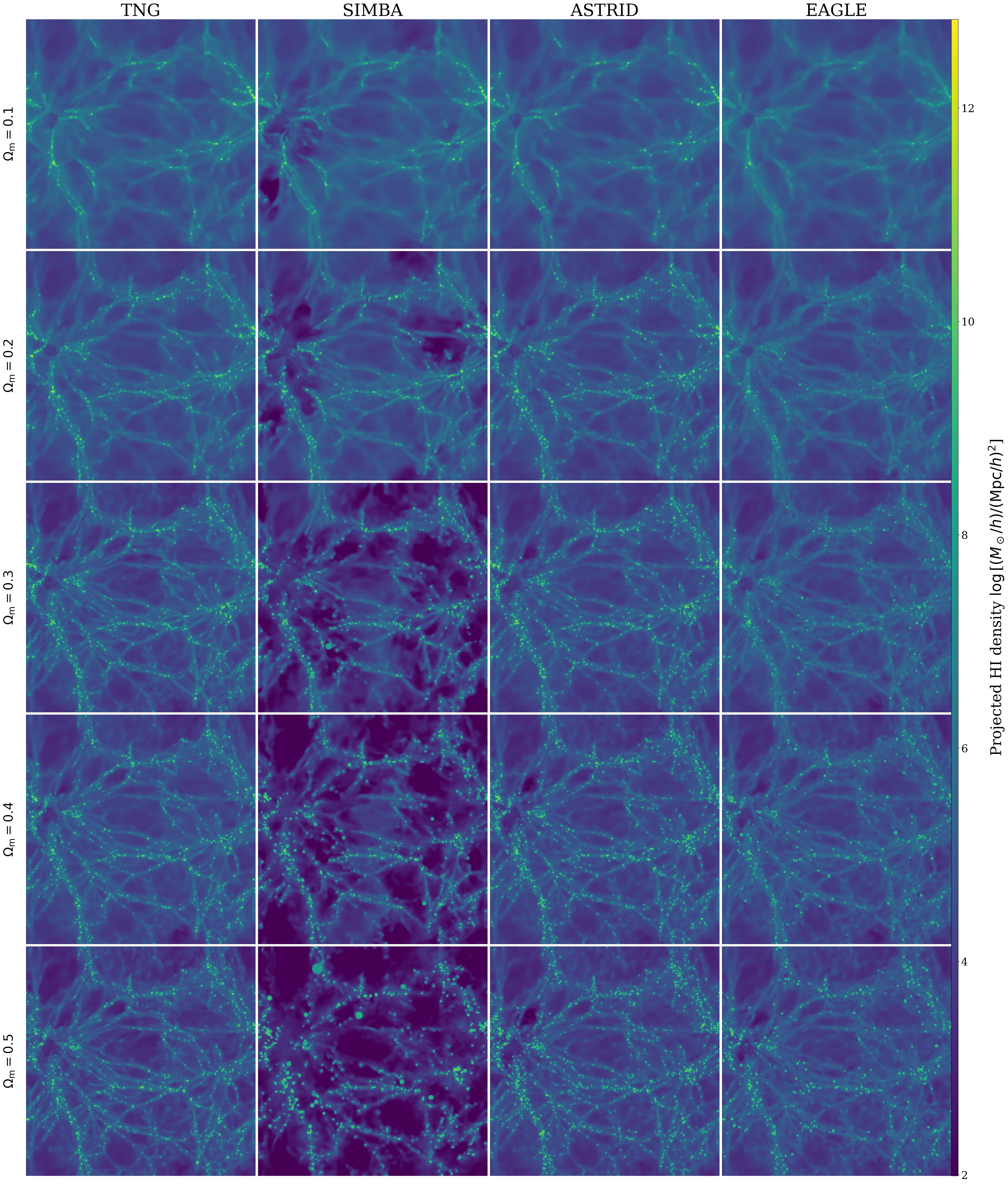}
    \caption{HI maps from \tng{}, \simba{}, \astrid{}, and \eagle{} with variation of $\om$ with the other parameters fixed.
    The initial conditions of the simulations are designed to yield the same large-scale structure across different simulations. 
    $\om$ has been set to 0.1 ({\it 1st row}), 0.2 ({\it 2nd row}), 0.3 ({\it 3rd row}), 0.4 ({\it 4th row}), and 0.5 ({\it 5th row}).
    The simulation run with $\om=0.3$ ({\it middle}) is the fiducial CAMELS simulation.}
    \label{fig:hi_maps_om}
\end{figure*}

\begin{figure*}
    \includegraphics[width=0.98\linewidth]{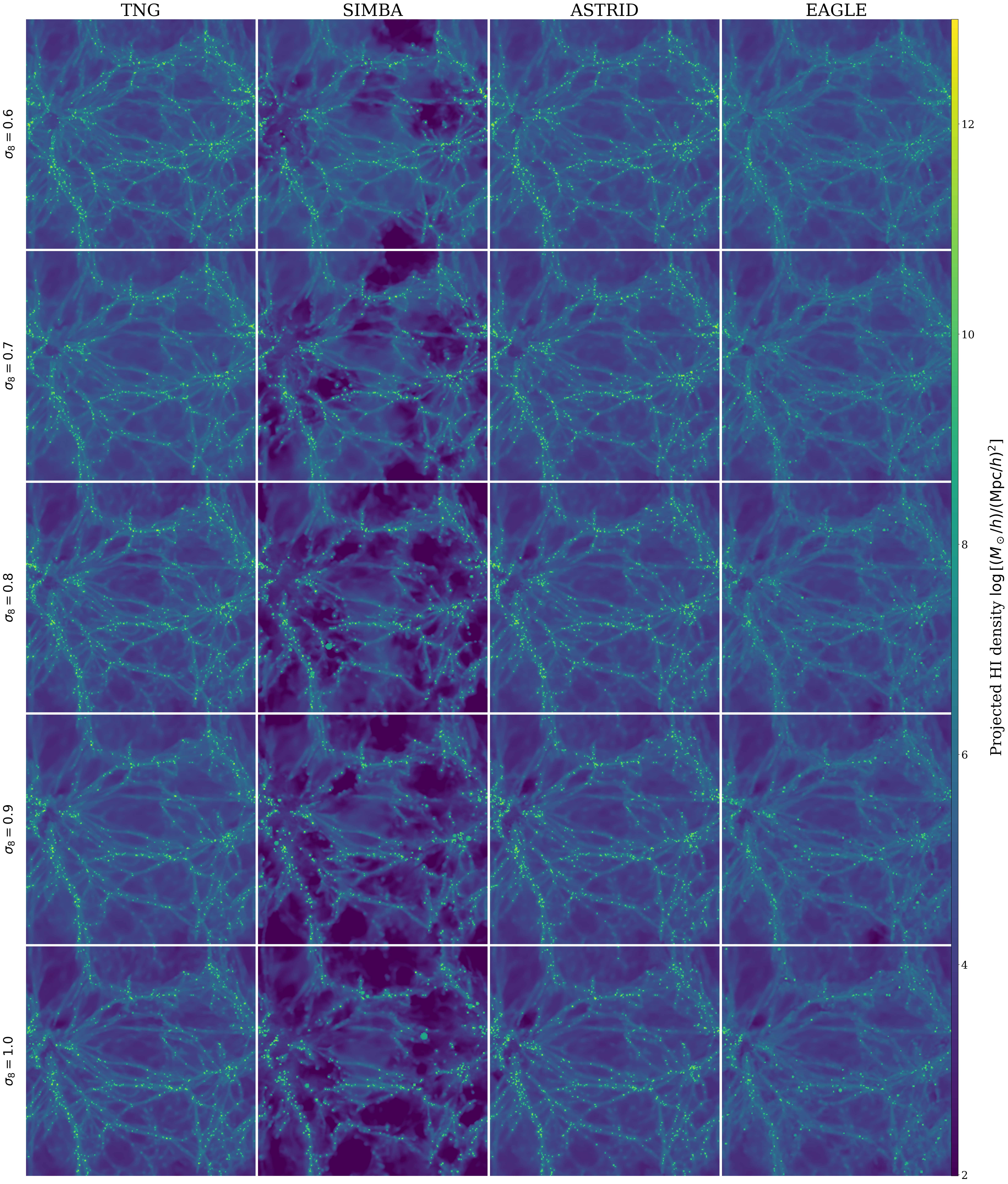}
    \caption{HI maps from \tng{}, \simba{}, \astrid{}, and \eagle{} with variation of $\sig$ with the other parameters fixed.
    The initial conditions of the simulations are designed to yield the same large-scale structure across different simulations. 
    $\sig$ has been set to 0.6 ({\it 1st row}), 0.7 ({\it 2nd row}), 0.8 ({\it 3rd row}), 0.9 ({\it 4th row}), and 1.0 ({\it 5th row}).
    The simulation run with $\sig=0.6$ ({\it middle}) is the fiducial CAMELS simulation.}
    \label{fig:hi_maps_sig}
\end{figure*}

\subsection{Different Combinations Of Simulations}
\label{sec:discussion_difference_in_simulations}
Thus far, we have conducted this study using the machine only trained on \tng{} and \simba{}. 
In this section, we extend our scope to all four simulations and investigate differences between all possible pairs of simulations. 
To this end, we use three approaches;
1. We train a {\miest} machine on two simulations without robustness to assess the degree to which the two simulations are distinguishable;
2. We individually train CNNs on each simulation and evaluate their performances on the other three simulations to obtain a comprehensive understanding of cross-simulation performance;
3. We perform a comparative analysis of the HI maps derived from the four simulations based on the results above.


First, we train \miest{} without robustness on every combination of any two simulations out of the four simulations---\tng{}, \simba{}, \astrid{}, and \eagle{}, which gives us six combinations in total.
Due to lack of robustness, the \miest{} is trained in a way that both compresses the maps and increases the mutual information between latent variables and cosmological parameters.
Since the de-classification is not imposed, there is no need for \miest{} to remove the classifiable information.
For instance, this is the case in Sec. \ref{sec:latent_space} where \miest{} without robustness shows that the latent variables from \tng{} and \simba{} are apparently classifiable.
Thus, we then adopt a evidence probability of model to measure the distinguishability from a binary classifier.
In this process, we train around 100 \miest{} for each combination and calculate the mean and the standard deviation to get rid of the randomness in the result.
Secondly, we individually train simple CNNs on each simulation and then evaluate the cross-performances by testing each CNN on the three simulations other than the one used for training. 
To mitigate the effects of randomness, we again repeat it over 100 times, and obtain the mean and the standard deviation for each combination.

\begin{figure*}
    \centering
    \includegraphics[width=0.98\linewidth]{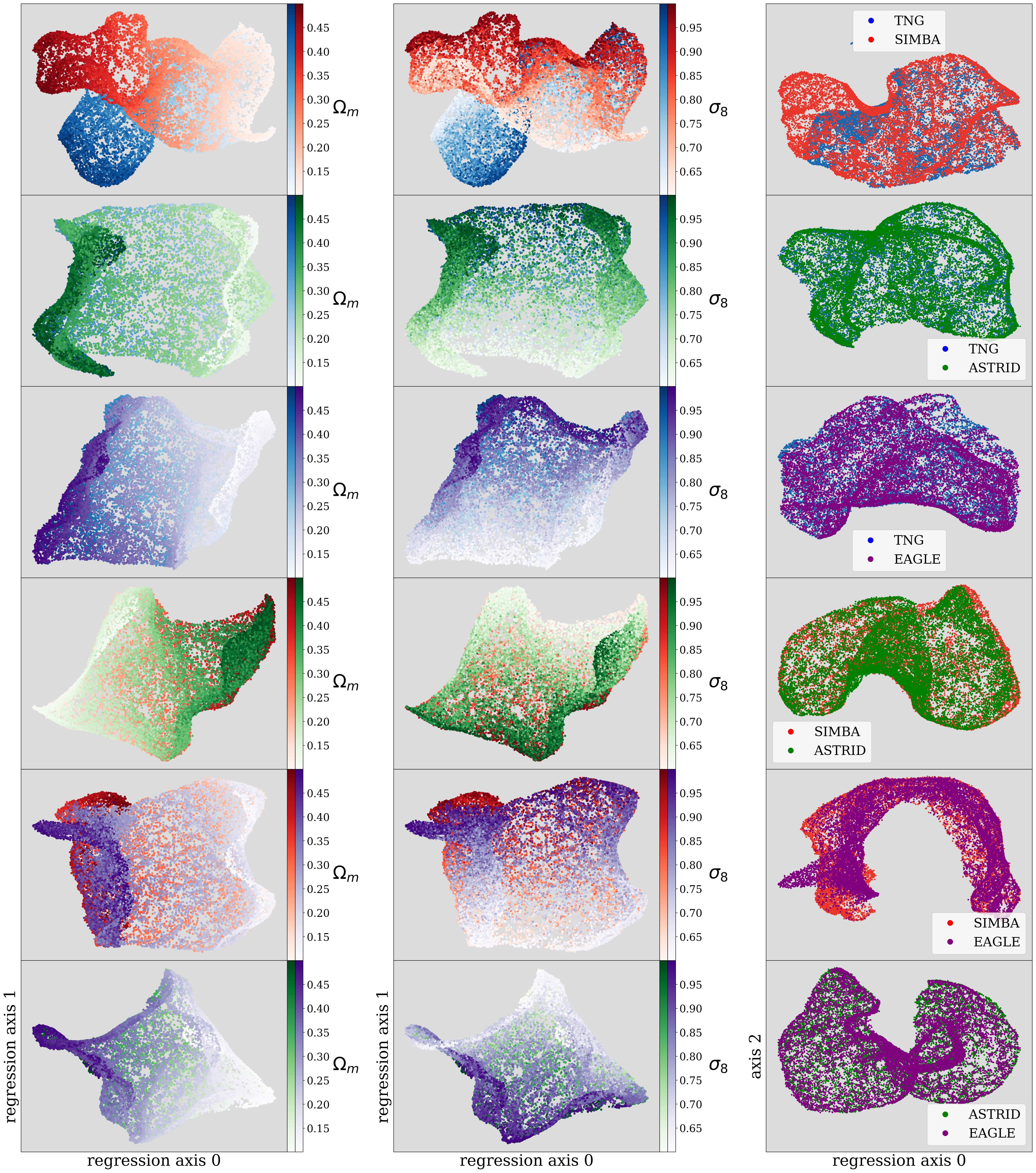}
    \caption{Latent space visualization using UMAP with the `canberra' metric from the six \miest{}s trained on all possible combinations of two simulations without robustness.
    The same machines used in generating Tab.~\ref{tab:two_simulations} are employed (see Sec.~\ref{sec:discussion_difference_in_simulations} for details).
    }
    \label{fig:simulations_full_combinatory}
\end{figure*}

Tab. \ref{tab:two_simulations} summarizes the probability of a simulation model $A$ given latent variables of a certain model--- $Z_B$, $P(A|Z_B)$, and AUC score {\it without} robustness.
$P(A|Z_B)$ represents the probability of \miest{} that is trained on Simulation A and Simulation B identifying a given input as either one of those simulations.
Here, we present the results for all six combinations of the four simulations.
Tab. \ref{tab:two_simulations} remarkably exhibits consistent results in terms of the relationships between the four simulations.
Specifically, \tng{}, \astrid{} and \eagle{} share more common features, whereas \simba{} is an outlier.
All combinations of two simulations amongst \tng{}, \astrid{}, and \eagle{} show probabilities $P(B|Z_A)$ of $\sim0.5$ and AUC scores of $\sim0.6$, while all combinations that include \simba{} exhibit probabilities $P(B|Z_A)$ of $\sim0.3$ and AUC scores of $\sim0.8$ (c.f., Fig. \ref{fig:simulations_full_combinatory} for an extensive latent space comparison of all four simulations).

One interesting observation is that despite the consistency of these trends, there exists a notable asymmetry.
For instance, $P(\mathrm{\tng}|Z_\mathrm{\astrid})=0.49$ while $P(\mathrm{\astrid}|Z_\mathrm{\tng})=0.45$, and $P(\mathrm{\simba}|Z_\mathrm{\astrid})=0.34$ while $P(\mathrm{\astrid}|Z_\mathrm{\simba})=0.30$.
Despite the comparable sizes of the variances ($\sim0.02$ for the \tng{}-\astrid{} pair and $\sim0.06$ for the \simba{}-\astrid{} pair), the presence of these asymmetric features suggests that certain simulations are relatively more distinctive, making them more challenging to blend effectively.
In this spirit, $\sum_{A}P(A|Z_\mathrm{B})$ can be a proxy for measuring how general or agreeable the simulation B is with other simulations. 
Then, $\sum_{A}P(A|Z_\mathrm{\astrid})$ shows the highest value of $0.43$, whereas
$\sum_{A}P(A|Z_\mathrm{\simba})$ exhibits the lower value of $0.29$.
In other words, this implies that \astrid{} possesses more common features and is more agreeable, while \simba{} stands out as the most distinctive simulation model.

In consideration of this, we proceed to analyze the projected HI maps to gain a direct insight into the analysis above.
Figs. \ref{fig:hi_maps_om} and  \ref{fig:hi_maps_sig} display projections of the HI maps from the four simulations with variations of cosmological parameters.
The initial conditions of the simulations are designed to yield the same large-scale structure across different simulations. 
Fig. \ref{fig:hi_maps_om} depicts the HI maps with different values of $\om$ with the other parameters fixed---$\om=[0.1,0.2,0.3,0.4,0.5]$ from the {\it top}, while Fig. \ref{fig:hi_maps_sig} illustrates the HI maps with different values of $\sig$ with the other parameters fixed---$\sig=[0.6,0.7,0.8,0.9,1.0]$ from the {\it top}.

The large-scale structure of all the presented HI maps---commonly known to be determined by physics of dark matter and the initial conditions of cosmological simulations---shows a great similarity across different simulations.
However, appreciable differences in the general features are observable with ease. 
The primary distinction is the amount of ionized diffuse gas.
In every map presented here, it is readily apparent that \simba{} contains a significantly greater amount of ionized diffuse gas---dark blue regions---than the other simulations, with this disparity becoming more prominent as $\om$ and $\sig$ increase. 
This phenomenon is also completely in agreement with Fig. \ref{fig:latent_tng_simba} that demonstrates significantly increasing distances between \tng{} and \simba{} in the latent space \tr{$\tilde{\mathcal{Z}}_\mathrm{UMAP}$} as $\om$ and $\sig$ increase.
On the other hand, \tng{}, \astrid{}, and \eagle{} exhibit similarity in terms of hot diffuse gas. 
This similarity accounts for the indistinguishability amongst \tng{}, \astrid{}, and \eagle{}, while also elucidating the distinguishability with \simba{}, shown in Tab. \ref{tab:two_simulations}.

This disparity is also demonstrated in \citet{tillman2023AJ....166..228T} that has studied the impact of galactic feedback on Ly$\alpha$ forest statistics, which is characterized by column densities of HI, resulting from parameter variations in the \simba{} and \tng{} feedback models. 
The study highlights that in \simba{}, the AGN jet feedback mode effectively transports energy to the diffuse IGM, and stellar feedback is crucial in regulating supermassive black hole growth and feedback, emphasizing the need to constrain both stellar and AGN feedback. 
In contrast, the \tng{} simulations show that AGN feedback variations do not influence the Ly$\alpha$ forest, but changes in the stellar feedback model lead to subtle effects \citep[see][for a further discussion on \simba{}]{tillman2024arXiv241005383T}.
This result elucidates the intrinsic discrepancies in the galactic feedback models in \tng{} and \simba{}, which results in the distinguishability.

On the other hand, the indistinguishability---e.g., amongst \tng{}, \astrid{}, and \eagle{}---does {\it not} necessarily result in better cross-performance.
It is worth noting that the cross-performance depends on mutual information between the cosmological parameters and the HI maps, whereas classification occurs on each map.
Thus, the cross-performance is sensitive to how structures, such as the clustering of galaxies, changes with respect to the cosmological parameters, whose relationship is not readily discernible to the naked eye.
For instance, \eagle{} in comparison to \tng{} and \astrid{} displays smoother structures, while \tng{} and \astrid{} exhibit denser clumping, albeit this density is not always pronounced.
In addition, such differences are not solely driven by the cosmology but also driven by the baryon physics even without varying the subgrid model parameters.
For example, increasing the matter content in the universe (high $\om$) results in the formation of dense clusters and galaxies, which facilitates the growth of supermassive black holes and subsequently leads to a greater fraction of hot diffuse gas in the universe due to the jet feedback from these black holes.
Consequently, this can produce effects comparable to the increase in the AGN parameters. 
In other words, the estimation of cosmological parameters can be significantly influenced by the subgrid models utilized in cosmological simulations. 
This underscores the necessity for robustness in these simulation models.

\section{Summary}
\label{sec:summary}
In this work, we develop the \miest{} (model-insensitive estimator) that can robustly estimate the cosmological parameters $\om$ and $\sig$ from HI maps of various cosmological simulation models in the CAMELS simulations.
The cosmological simulation models employed in this study include \tng{}, \simba{}, \astrid{}, and \eagle{}.
Our primary objective is to train \miest{} on \tng{} and \simba{} {\it with} and {\it without} robustness, followed by testing on \astrid{} and \eagle{} to evaluate the robustness of the model.
Here, an estimator is considered {\it robust} if it posseses a consistent predictive power across all simulations, including those used during the training phase.

To achieve robustness, we construct the \miest{} by integrating the approaches of deep variational information bottleneck \citep{alemi2017deep} and generative adversarial network \citep{ganNIPS2014_5ca3e9b1}. 
The structure of \miest{} is comprised of a CNN, an encoder, a regressor, and a classifier.
The CNN and an encoder image-process and compress the HI maps into latent variables, which can be deemed as summary statistics of these maps.
With the latent variables, the \miest{} can predict cosmological parameters via the regressor.
During the training phase, the \miest{} aims to discard the classifiable information specific to each simulation through de-classfication process with the classifier, while preserving the information about the cosmological parameters (refer to Sec. \ref{sec:method_neural_network} for details).
Subsequently, the hyperparameters and training procedures are optimized using {\tt optuna} to improve the performance of the machine.

We first investigate the latent space---a set of summary statistics within neural networks that captures the essential patterns or features of complex data---of the \miest{} to understand the mechanism through which \miest{} attains robustness and to assess the efficacy of the \miest{} (see Sec.~\ref{sec:latent_space}).
We utilize methods for reducing dimensionality that provide meaningful evaluations of high-dimensional data, condensing the latent space to three dimensions.
On the three dimensional latent space derived from the test set of \tng{} and \simba{}, \miest{} without robustness  generates two separate sheets of latent variables originating from \tng{} and \simba{}, respectively.  
The two axes of the sheets correspond to $\om$ and $\sig$. 
However, the introduction of robustness results in these two sheets transforming into two intermingled spheres, showing no sign of separation or distinctiveness, while preserving the two axes for $\om$ and $\sig$.
Consequently, the removal of the classifiable information about simulation models leads to the blending of latent variables.

We turn to examine how robustness facilitates the incorporation of the training simulations with the unseen simulations.
To this end, we focus on the overlapping area in the latent space between the training simulations---\tng{} and \simba{}---and the unseen simulations---\astrid{} and \eagle{}.
The implementation of robustness increases coverage of \astrid{} and \eagle{} on \tng{} and \simba, leading to improved alignment of the axes for $\om$ and $\sig$.
This potentially indicates better performance on \astrid{} and \eagle{}.
Nonetheless, biases persist, particularly in specific regions such as high $\om$ areas for \eagle{} with robustness, suggesting that systematic biases and inaccuracies still exist despite the improvements. 
Consequently, the incorporation of robustness helps align training and unseen simulations, but some persistent biases are likely due to the inherent properties of the simulations.

Subsequent to the qualitative analysis, we transition to a quantitative assessment of the robustness and performance for \miest{} (refer to Sec. \ref{sec:performance}).
\tr{For an estimator trained on IllustrisTNG and SIMBA without robustness}, the mean relative errors of $\om$ and $\sig$ are high for \astrid{} ($\sim$17\%) and \eagle{} ($\sim$15\%) compared to \tng{} ($\sim$ 4\%) and \simba{} ($\sim$5\%), showing clear bias. 
Incorporating robustness improves performance of \astrid{} and \eagle{} considerably, reducing the mean relative errors of $\om$ and $\sig$ by $\sim$35\% (\astrid) and $\sim$20\% (\eagle) and decreasing biases by factors of 7 (\astrid) and 4 (\eagle). 
These improvements, attributed to enhanced coverage and alignment in the latent space, underscore the importance of robustness in improving consistent performance over various simulations.

In addition to the enhancement in performance, two observations can be highlighted. 
Firstly, the improvement is more significant in $\sig$ compared to $\om$. Without robustness, the accuracy of $\sig$ in both \astrid{} and \eagle{} exhibits considerable bias, yielding values between 0.8 and 1.0 for $\sig$, regardless of the actual true value. 
Nonetheless, the incorporation of robustness significantly alleviates these biases, showing a higher correlation with the true values. 
Conversely, the improvement in $\om$ is relatively minor compared to $\sig$. 
The second is that the prediction of $\om$ in \eagle{} is significantly biased regardless of the presence of robustness.
This underscores the inherent divergence in the physical models, as shown in the latent space that illustrates the substantial absence of data points in the high $\om$ region both with and without robustness.

We then investigate the changes of performance of \miest{} in response to the different degrees of robustness.
The incorporation of robustness increases the errors for \tng{} and \simba\ but improves the accuracy for \astrid{} and \eagle. 
Specifically, the accuracy of $\sig$ is more significantly improved by the introduction of robustness than $\om$. 
Furthermore, the performance variations with different robustness levels are analyzed using hyperparameters $\beta$ and $\gamma$ in Eq.~\ref{eq:loss_function_1} in Sec.~\ref{sec:method_neural_network_equations}, which control the degrees of compression and de-classification. 
For the training simulations, increased robustness leads to reduced predictive power due to information loss from compression and de-classification, respectively. 
In the unseen simulations, the accuracy gained through robustness shows an initial increase, followed by a decline after reaching an optimal point of maximum robustness.
The turning point for $\om$ occurs earlier than for $\sig$, indicating that $\om$ is more sensitive to compression and de-classification.



\acknowledgments
The Flatiron Institute is supported by the Simons Foundation.
This work was supported by the Simons Collaboration on “Learning the Universe”.

\bibliography{main}{}
\bibliographystyle{aasjournal}

\end{document}